\documentclass[aps,pre,reprint,superscriptaddress,longbibliography,groupedaddress]{revtex4-2}%nofootinbib

\usepackage{graphicx} % Required for inserting images
\usepackage{amsmath}
\usepackage{bm}
\usepackage{amssymb}
\usepackage{dsfont}

\usepackage{verbatim}
\usepackage[T1]{fontenc}
\usepackage[english]{babel}
\usepackage[utf8]{inputenc}

\usepackage[colorlinks = true, citecolor = black, linkcolor = black, urlcolor=black]{hyperref}

\begin{document}
	
	\title{Occupation Uncertainty Relations}
	\author{Katarzyna Macieszczak}
	\affiliation{Department of Physics, University of Warwick, Coventry CV4 7AL, United Kingdom}

\begin{abstract} 
	We introduce occupation uncertainty relations (OURs) for dynamics of a Markov process over discrete configurations.  Those are lower bounds on uncertainties of system observables that are time-integrated along stochastic trajectories.   {The uncertainty is defined as the ratio of the variance to the square of the average, with the latter necessarily shifted by the average in a reference distribution. The derived bounds are observable independent, but rely on the reference choice. We show, however, that all OURs originate from a matrix bound on correlations between times spent in different system configurations, i.e., occupation times. This result is reference independent, and expressed only in terms of stationary probability and lifetimes of individual configurations. Any dynamics that saturates the matrix bound, is optimal in approximating stationary distribution by occupation times; this always occurs for unidirectional cycles. Furthermore, using this bound, dynamic correlations can be asymptotically bounded in terms of static correlations, in turn leading to the positivity of dynamic exponents in the thermodynamic limit. While  at a finite size the matrix bound generally saturates away from equilibrium, we demonstrate detailed-balance dynamics can still saturate it asymptotically, by leading to vanishing dynamic exponents and optimal simulations in that context}.
\end{abstract}

\maketitle

\emph{Introduction}.
For mesoscopic systems, stochastic changes in their configurations lead to non-negligible fluctuations of thermodynamics quantities~\cite{Seifert2012}. For fluxes and currents those have been shown to be nevertheless limited in terms of so called thermodynamic~\cite{Barato2015,Gingrich2016,Dechant2018,Dechant2018a,Ito2020,Liu2020,Koyuk2020} and kinetic~\cite{Garrahan2017,Terlizzi2019,Koyuk2020} uncertainty relations (TUR and KUR).  
The striking feature of these relations is that the bounds, which hold from below, are given by time integrals of local-in-time thermodynamic properties -- the entropy production rate and the activity, respectively -- and thus they admit long-time limits. This is in apparent contrast with another class of uncertainty relations, so called generalised TUR~\cite{Hasegawa2019,Ziyin2023}, with bounds in terms of properties of stochastic trajectories as a whole.

Importantly, the TUR and the KUR hold not only for currents and fluxes, but can involve system observables that however enter the uncertainties in a modified way~\cite{Koyuk2020}. Instead of dividing the variance of the time integral by the squared average at a final time, as done for currents and fluxes,  that average is additionally shifted, {e.g., in the case of time-independent dynamics, by the average of the scaled time integral, so that the average in the ratio is replaced by the difference between those averages.
	As a consequence, the TUR and the KUR become trivial when used for system observables at long times (or when initial configurations are sampled from the stationary distribution~\cite{Horowitz2017}). Indeed, in that case those averages become the same, and their vanishing difference causes the uncertainty of any system observables to diverge while the bounds -- the entropy production rate and the activity -- remain finite. 
	
	In this work, we rectify this situation by considering uncertainties for system observables featuring a difference to their average with respect  to an arbitrary reference distribution, which in general no longer diverges in the asymptotic limit. 
	For those, we derive observable-independent lower bounds, which gives rise to non-trivial OURs. We proceed to show that such relations stem from a lower matrix bound on correlations between occupation times. For a given observable, it reduces to a faithful lower bound on the variance, which attains $0$ only if the observable is constant.  The matrix bound,  expressed in terms of the stationary distribution and lifetimes of individual configurations, is independent from a reference choice, constituting a fundamental limit on rates of occupation time fluctuations for any Markov dynamics. From a practical perspective, it sets an error benchmark for any stochastic simulations using occupation times to approximate stationary averages of system observables. 
	We show that the bound always saturates for unidirectional cycles and, more generally, if any Markov process also satisfies this property is away from equilibrium. In the context of non-equilibrium statistical physics, this bound allows us to connect the dynamic and static correlations, and to consider well-defined dynamic exponents~\cite{Sokal1997}, e.g., near critical points. 	
	Interestingly, while equilibrium dynamics cannot saturate the bound at finite sizes, it is still possible for it to yield zero dynamic exponents  in the thermodynamic limit.}\\

\begin{figure*}[t]
	\includegraphics[width=1.0\linewidth]{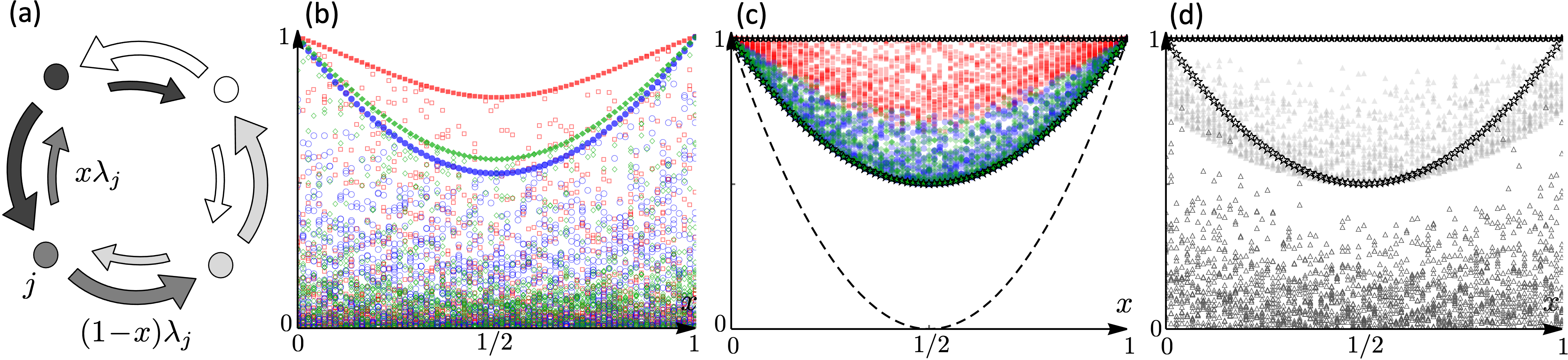}
	\caption{
		\label{fig:Markov}\textbf{OUR and mOUR for Markov cycles}. 
		\textbf{(a)} The dynamics for $d=4$ configurations {in \emph{Example I}}.
		\textbf{(b)}~The~quality of the OUR in  Eq.~\eqref{eq:OUR} with random references (empty markers; $\mathbf{p}$ uniformly sampled for each $x$) and optimal references {[filled markers; coincides with  the quality of Eq.~\eqref{eq:vOUR}]}, for $(\mathbf{o})_j=\cos(j \pi)$, $\cos(j 2\pi/d)$, $\sin(j\pi/2)$ (red squares, blue circles, green diamonds) and $\lambda_j\propto 2^j$.
		\textbf{(c)} 
		The best and worst performance  of the mOUR in Eq.~\eqref{eq:mOUR} (filled and empty black stars)
		and the quality of Eq.~\eqref{eq:vOUR} for the observables in (a) (filled markers)  for random cycles [$\mathbf{p}_\text{ss}$  uniformly sampled for each $x$; {$\lambda_j\propto 1/(\mathbf{p}_\text{ss})_j$}]; when $d\rightarrow\infty$, the best performance of the mOUR for uniform cycles is $1$, the worst shown as black dashed line {($\lambda_j \propto 1$)}.
		\textbf{(d)}~The best and worst performance  of the bound in Eq.~\eqref{eq:mOUR_var} for random cycles (filled and empty grey triangles; $\mathbf{p}_\text{ss}$ uniformly sampled for each $x$) and uniform cycles [filled and empty black stars; as in (c)].
	} 
\end{figure*}

\emph{Uncertainties and References}. {We begin by explaining the necessity of reference distributions when considering uncertainties of system observables}.  The uncertainty of  fluxes and currents is defined as the ratio of their (time-integrated) variance to the square their (final) average. 
This renders the uncertainty invariant under their rescaling by a (multiplicative) constant, $\alpha$,  since the average rescales by $\alpha$ and the variance by $\alpha^2$. This holds analogously for system observables.
Unlike observables for fluxes and currents, however, which depend on the number of stochastic transitions between configurations and vanish when no transition occurs, 
system observables take values determined by the currently occupied configuration, $O(t)\equiv\sum_{j} \delta_{j}{(t)}\,o_{j}$ with $\delta_{j}{(t)}=1$ if $j$ occupied and $0$ otherwise. Since some configuration is occupied at any time, $\sum_j \delta_{j}(t)=1$, system observables can also be shifted by an (additive) constant, $\beta$, with $o_j \mapsto  o_j+\beta$ leading to $O(t)\mapsto O(t)+\beta$, and for their time integrals to, $\overline{O}(t)\equiv \int_0^t dt'\, O(t') \mapsto  \overline{O}(t)+ t\beta$. As the averages $\langle O(t) \rangle$, $\langle \overline{O}(t) \rangle$ shift analogously, while the variances $\Delta^2 O(t) $, $\Delta^2 \overline{O}(t)$ remain the same, the  uncertainty given by the ratio of  the variance to the squared average, which we will refer to the \emph{standard uncertainty}, vanishes $\propto1/\beta^{2}$ when $\beta\rightarrow \pm\infty$.  
Thus, for a non-zero lower limit on such uncertainties to exist, constant contributions need to be removed.

Let $(\mathbf{o})_j\equiv o_j$ and $(\mathbf{1})_j\equiv 1$. For a projection $\mathbf{Q}=\mathbf{Q}^2$ with the kernel containing $\mathbf{1}$, projected observables are left  invariant by constant shifts, $\mathbf{Q}\mathbf{o}=\mathbf{Q}(\mathbf{o}+\beta \mathbf{1})$. Since $\langle O(t)\rangle=\mathbf{p}(t)^\text{T}\mathbf{o}$, where $[\mathbf{p}(t)]_j\equiv\langle\delta_{ j }(t)\rangle$ is the probability of the system found  in a configuration $j$ at time $t$, it follows that the averages (and higher cumulants) of projected observables remain unchanged by constant shifts, and so do their time integrals. A  highest dimensionality projection image corresponds to the single-dimensional kernel, so that $\mathbf{Q}=\mathbf{I}-\mathbf{1}\mathbf{p}^\text{T}$, where $(\mathbf{I})_{jk}\equiv\delta_{jk}$ and $\mathbf{p}^\text{T}\mathbf{1}=1$. Then, the variances (and higher cumulants) for observables and their projections are exactly the same, and this holds for their time integrals too, so we will consider only such projections from now on.

When $(\mathbf{p})_j\geq 0$ for all $j$,  $\mathbf{p}$ takes the role of a reference probability distribution. Indeed,  the average of a projected observable equals the difference between its average before the projection, and such an average for $\mathbf{p}${, e.g.,  $\mathbf{p}(t)^\text{T}(\mathbf{Q}\mathbf{o})=\mathbf{p}(t)^\text{T}\mathbf{o}-\mathbf{p}^\text{T}\mathbf{o}$}. We will refer to its square dividing the variance in the corresponding uncertainty as the \emph{relative uncertainty}. 
We show below that this relationship between the standard uncertainty of the projected observables and the relative uncertainty of the observable holds more generally, but up to a multiplicative constant, provided that the probability distribution determining the  average before the projection
is supported everywhere. \\

\emph{OURs}. For a Markov process with a time-independent generator $\mathbf{W}$ [with $(\mathbf{W})_{jk}\geq 0$ for $j\neq k$ and  $\sum_{k}(\mathbf{W})_{jk}=0$, see, e.g.,~Fig.~\ref{fig:Markov}(a)],  $d\mathbf{p}(t)^\text{T}/dt= \mathbf{p}(t)^\text{T} \mathbf{W}$.  Let $\mathbf{p}_\text{ss}$ denote a stationary distribution [$ \mathbf{p}_\text{ss}^\text{T}\mathbf{W}=\mathbf{0}^\text{T}$]. Assuming the ergodic dynamics,  $\mathbf{p}_\text{ss}$ is unique, so that $\lim_{t\rightarrow \infty} \mathbf{p}(t)= \mathbf{p}_\text{ss}$ for any initial condition, and $(\mathbf{p}_\text{ss})_j>0$. While the averages of system observables and their time-integrals are asymptotically determined by the stationary distribution, $\lim_{t\rightarrow \infty} \langle O(t)\rangle=\lim_{t\rightarrow \infty} \langle \overline{O}(t)\rangle /t= \langle O\rangle_\text{ss}\equiv
\mathbf{p}_\text{ss}^\text{T}\mathbf{o}$, 
for the variances this holds true only for observables at  a final time, $\lim_{t\rightarrow \infty}\Delta^2 O(t)= \mathbf{o}^\text{T}\mathbf{\Delta}^2\mathbf{o}$ with $\mathbf{\Delta}^2\equiv\mathbf{D}- \mathbf{p}_\text{ss}\mathbf{p}_\text{ss}^\text{T}$ and $(\mathbf{D})_{jk}=\delta_{jk}(\mathbf{p}_\text{ss})_j$. The asymptotic rates of the variances for the time integrals
depend also on the generator via the resolvent $\mathbf{R}$ as $\lim_{t\rightarrow \infty} \Delta^2 \overline{O}(t)/t= \mathbf{o}^\text{T}\mathbf{C}\mathbf{o}$ with $\mathbf{C}\equiv - (\mathbf{D}\mathbf{R}+\mathbf{R}^\text{T}\mathbf{D})$ [where  $\mathbf{R}\mathbf{W}=\mathbf{W}\mathbf{R}= \mathbf{Q}_\text{ss}\equiv \mathbf{I}-\mathbf{1}\mathbf{p}_\text{ss}^\text{T}$].

Let $\langle O\rangle\equiv\mathbf{p}^\text{T}\mathbf{o}$ be the average for a reference distribution $\mathbf{p}$. Exploiting the large deviation theory~\cite{denHollander2000,Touchette2009}, namely 
the level-2.5 rate function for occupations and fluxes in Markov processes~\cite{Maes2008,Bertini2015},  we show in~\cite{SM} that the asymptotic rate of the relative uncertainty is bounded from below in terms of the difference between the stationary and reference distributions as
\begin{align}\notag
	\lim_{t\,\rightarrow \infty}& \,\frac{1}{t}\,\frac{\Delta^2 \overline{O}(t)}{[\langle \overline{O}(t)\rangle/t -\langle O\rangle]^2}
	=\frac{\mathbf{o}^\text{T} \mathbf{C}  \mathbf{o}}{[(\mathbf{p}_\text{ss}-\mathbf{p})^\text{T}\mathbf{o} ]^2}\\
	\label{eq:OUR}
	&\geq \frac{1} {(\mathbf{p}_\text{ss}-\mathbf{p})^\text{T} \mathbf{B}  (\mathbf{p}_\text{ss}-\mathbf{p})-\frac{[(\mathbf{p}_\text{ss}-\mathbf{p})^\text{T}\mathbf{\Lambda}\mathbf{1}]^2}{\mathbf{p}_\text{ss}^\text{T}\mathbf{\Lambda}\mathbf{1}}}.
\end{align}
Here, 
$\mathbf{B}\equiv\mathbf{D}^{-1}\mathbf{\Lambda}=\mathbf{\Lambda}\mathbf{D}^{-1}$ and $(\mathbf{\Lambda})_{jk}\equiv-\delta_{jk}(\mathbf{W})_{jj}$ encodes the decay rates, so that $\mathbf{p}_\text{ss}^\text{T}\mathbf{\Lambda}\mathbf{1}>0$ is the activity.  Furthermore, the rescaled average of the time-integral, $\langle \overline{O}(t)\rangle/t$, can be replaced by the average at final time, $\langle O(t)\rangle $ since their limit is the same.

When the reference distribution approaches the stationary distribution, $\mathbf{p}=\mathbf{p}_\text{ss}+\epsilon\boldsymbol{\delta}\mathbf{p}$ with $\epsilon\rightarrow 0$ (and $\boldsymbol{\delta}\mathbf{p}^\text{T}\mathbf{1}=0$),  the relation is \emph{tight}, with both the bound and  the uncertainty vanishing as $\epsilon^2$.
Actually, by considering the ratio of  the bound to the uncertainty,  which we define as the bound \emph{quality},  we can again obtain Eq.~\eqref{eq:OUR} in the limit $\epsilon\rightarrow 0$ but now with $\mathbf{p}-\mathbf{p}_\text{ss}$ replaced by $\boldsymbol{\delta}\mathbf{p}$. 
Redefining $\mathbf{p}=\mathbf{p}_\text{ss}+\boldsymbol{\delta}\mathbf{p}$, we then find Eq.~\eqref{eq:OUR} holds for any $\mathbf{p}$ such that $\mathbf{p}^\text{T}\mathbf{1}=1$. Since $(\mathbf{p}_\text{ss}-\mathbf{p})^\text{T} =\mathbf{p}_\text{ss}^\text{T}\mathbf{Q}$, so that $(\mathbf{p}_\text{ss}-\mathbf{p})^\text{T}\mathbf{o} =\mathbf{p}_\text{ss}^\text{T}  (\mathbf{Q}\mathbf{o}) $, while $\mathbf{o}^\text{T} \mathbf{C}  \mathbf{o}=(\mathbf{Q}\mathbf{o})^\text{T}  \mathbf{C} (\mathbf{Q}\mathbf{o})$ Eq.~\eqref{eq:OUR} can be then be equivalently viewed as the bound $1/[\mathbf{p}_\text{ss}^\text{T} \mathbf{Q}\mathbf{B} \mathbf{Q}^\text{T}\mathbf{p}_\text{ss} -(\mathbf{p}_\text{ss}^\text{T} \mathbf{Q}\mathbf{\Lambda} \mathbf{1})^2/(\mathbf{p}_\text{ss}^\text{T}\mathbf{\Lambda}\mathbf{1})]$, for the standard uncertainty of time-integrals of projected observables. Thus,  we will simply refer to that quantity as the uncertainty in the rest of this work.

While tighter bounds can also be found by exploiting level-2.5 rate function for Markov processes limited to occupations and currents~\cite{Bertini2015a,Barato2015b},  which we do in~\cite{SM}, they still hold non-trivially because of constant contribution eliminated by a method of choice.  The OURs lend themselves to eliminating this dependence all together as we show next. \\

\emph{Matrix OUR (mOUR)}.
Both the uncertainty and the bound in  Eq.~\eqref{eq:OUR} depend on the reference, or, equivalently, the projection [cf.~Fig.~\ref{fig:Markov}(b)]. 
Optimising the quality with respect to that choice, i.e., considering the best-case scenario for every observable, which is achieved with $\mathbf{p}=\mathbf{p}_\text{ss}+\epsilon \mathbf{Q}_\text{ss}^\text{T}\mathbf{B}^{-1} \mathbf{Q}_\text{ss} \mathbf{o}$,
we arrive at a fundamental result, 
\begin{align}\label{eq:vOUR}
	\lim_{t\rightarrow \infty} \,\frac{1}{t}\,\Delta^2 \overline{O}(t)
	%\frac{\Delta^2 \overline{O}(t)}{t}
	%&
	=\mathbf{o}^\text{T} \mathbf{C}  \mathbf{o}%\\
	%&
	\geq (\mathbf{Q}_\text{ss}\mathbf{o})^\text{T}\mathbf{B}^{-1}(\mathbf{Q}_\text{ss}\mathbf{o}),
\end{align}
where $\mathbf{B}^{-1}=\mathbf{D} \mathbf{\Lambda}^{-1}=\mathbf{\Lambda}^{-1}\mathbf{D}$, see~\cite{SM}.
This reference-independent bound on the asymptotic rate of the variance for time-integrated observables is invariant under both constant shifts of observables and their projections (recall that $\mathbf{Q}_\text{ss} \mathbf{1} =\mathbf{0}$ and note that $\mathbf{Q}_\text{ss}\mathbf{Q}=\mathbf{Q}_\text{ss}$).  Importantly, for finitely many configurations, unless the observable is constant, the bound is \emph{faithful}  (i.e., non-zero), cf.~Fig.~\ref{fig:Markov}(c). Indeed, $\mathbf{Q}_\text{ss}\mathbf{o}\neq \mathbf{0}$ unless $\mathbf{o}\propto \mathbf{1}$, while $\mathbf{B}^{-1}$ is a positive-definite matrix so that $(\mathbf{Q}_\text{ss}\mathbf{o})^\text{T}\mathbf{B}^{-1}(\mathbf{Q}_\text{ss}\mathbf{o})>0$.  
Furthermore, Eq.~\eqref{eq:vOUR} yields a better bound on the uncertainty than that in Eq.~\eqref{eq:OUR}.
Namely, $(\mathbf{Q}_\text{ss}\mathbf{o})^\text{T}\mathbf{B}^{-1}(\mathbf{Q}_\text{ss}\mathbf{o})/[\mathbf{p}^\text{T}(\mathbf{Q}_\text{ss}\mathbf{o})]^2$, 
where $[\mathbf{p}^\text{T}(\mathbf{Q}_\text{ss}\mathbf{o})]^2=[(\mathbf{p}_\text{ss}-\mathbf{p})^\text{T}\mathbf{o}]^2$. 
The OUR can be recovered from this bound as it corresponds to its worst-case scenario among all observables~\cite{SM},  $\min_\mathbf{o}(\mathbf{Q}_\text{ss}\mathbf{o})^\text{T}\mathbf{B}^{-1}(\mathbf{Q}_\text{ss}\mathbf{o})/[\mathbf{p}^\text{T}(\mathbf{Q}_\text{ss}\mathbf{o})]^2$, attained for 
$\mathbf{o}=\alpha\,\mathbf{B}[\mathbf{p}-\mathbf{p}_\text{ss}(\mathbf{p}^\text{T}\mathbf{\Lambda}\mathbf{1}) /(\mathbf{p}_\text{ss}^\text{T}\mathbf{\Lambda}\mathbf{1})]+\beta \mathbf{1}$ 
with $\alpha\neq 0$.

For observables with vanishing asymptotic averages, $\langle O\rangle_\text{ss}=0$, the bound is simply given in terms of $\mathbf{B}^{-1}$, which encodes the lifetimes of configurations (the inverse of their decay rate) weighted by their stationary probability.  In fact, Eq.~\eqref{eq:vOUR} can be equivalently viewed as an asymptotic bound on the correlation rates for occupation times~\footnote{We consider any inequalities between matrices according to Loewner order, i.e., $\mathbf{A}\geq \mathbf{B}$ if $\mathbf{A}-\mathbf{B}$ is positive semi-definite.},
\begin{equation}\label{eq:mOUR}
	\mathbf{C}\geq\mathbf{Q}_\text{ss}^\text{T} \mathbf{B}^{-1}\mathbf{Q}_\text{ss}.
\end{equation}
Indeed, $(\mathbf{C})_{jk}$ is the asymptotic rate for the correlation between total time spent in configurations $j$ and $k$, respectively [that is, the time-integrals of the observables with $o_i=\delta_{ij}$ and $o_i=\delta_{ik}$, respectively].  
Since their sum is fixed as equal to the final time $t$, the bounds on their variance rate depend on the global quantity  $\mathbf{p}_\text{ss}^\text{T}  \mathbf{\Lambda}^{-1}  \mathbf{1}$, i.e.,  $(\mathbf{C})_{jj}\geq\underline{} (\mathbf{p}_\text{ss})_j[1-2(\mathbf{p}_\text{ss})_j]/(\mathbf{\Lambda})_{jj} +(\mathbf{p}_\text{ss})_j^2 (\mathbf{p}_\text{ss}^\text{T}  \mathbf{\Lambda}^{-1}  \mathbf{1}) $~\footnote{While $(\mathbf{C})_{jj}\geq(\mathbf{p}_\text{ss})_j[1-2(\mathbf{p}_\text{ss})_j]/(\mathbf{\Lambda})_{jj}$ is a bound that is positive for $(\mathbf{p}_\text{ss})_j<1/2$, this bound cannot be saturated as the omitted non-local term is strictly positive.}. Similarly, for local observables this will be the case when using the bound in Eq.~\eqref{eq:vOUR}. \\

\emph{Saturating OURs and mOUR}. When Eq.~\eqref{eq:mOUR} saturates, $\mathbf{C}=\mathbf{Q}_\text{ss}^\text{T} \mathbf{B}^{-1}\mathbf{Q}_\text{ss}$, so does Eq.~\eqref{eq:vOUR} for any observable.  But the saturation of Eq.~\eqref{eq:mOUR} 
is not necessary for the saturation of Eq.~\eqref{eq:vOUR}, which can saturate for certain observables only.
The best/worst \emph{performance} of the bound in Eq.~\eqref{eq:vOUR}, i.e., the maximum/minimum ratio $\mathbf{o}^\text{T}\mathbf{Q}_\text{ss}^\text{T}\mathbf{B}^{-1}\mathbf{Q}_\text{ss}\mathbf{o}/(\mathbf{o}^\text{T} \mathbf{C}  \mathbf{o})$, is given by the maximum/minimum eigenvalue (on the orthogonal complement of $\mathbf{1}$) of $\sqrt{\mathbf{C}^+} \mathbf{Q}_\text{ss} ^\text{T}\mathbf{B}^{-1}\mathbf{Q}_\text{ss} \sqrt{\mathbf{C}^+}$ 
and achieved for observables given by corresponding eigenvectors multiplied by $\sqrt{\mathbf{C^+}}$,  $\mathbf{C}^+$ is the Moore-Penrose pseudoinverse  of $\mathbf{C}$ {(which features $0$ eigenvalue for $\mathbf{1}$ and only positive eigenvalues otherwise)}.  The saturation of Eq.~\eqref{eq:vOUR} is possible when the maximum equals $1$, while Eq.~\eqref{eq:mOUR} saturates when the minimum also equals $1$, cf.~Fig.~\ref{fig:Markov}(c). Similarly, the faithfulness of the bounds manifests as the minimum being strictly positive.

We now {explain how} the bound in Eq.~\eqref{eq:mOUR} can saturate only away from equilibrium, in contrast to the TUR~\cite{Barato2015,Macieszczak2018}, unless the structure of transitions with non-zero rates ensures the equilibrium (i.e., we exclude situations when no asymptotic currents can be present, e.g., in dynamics of $2$ configurations, but in~\cite{SM} 
we identify conditions for the saturation of Eqs.~\eqref{eq:OUR}-\eqref{eq:mOUR} -- analogously as done for the TUR in~Ref.~\cite{Shiraishi2021} -- and those cannot be satisfied for more configurations].
Later, we demonstrate that Eq.~\eqref{eq:mOUR} saturates for all unidirectional cycles (see also Fig.~\ref{fig:Markov}) of which dynamics of $2$ configurations is an example.

For ergodic dynamics, the asymptotic rate of the variance of a time-integrated observable  is known to be bounded from above by the corresponding rate in the detailed-balance dynamics with the identical stationary distribution and, in the long-time limit, the same average time-symmetric fluxes, but with zero average currents (see Refs.~\cite{Rey-Bellet2016,Kaiser2017} and, for Langevin dynamics, Refs.~\cite{Hwang2005,Duncan2016}).  Eq.~\eqref{eq:vOUR} can be viewed as a complementary lower bound, also identical for any dynamics associated with the same detailed-balance dynamics.
Moreover, the variance rate can generally be reduced among such dynamics by moving away from the equilibrium (through increasing the average currents, cf.~Lemma 2 in Ref.~\cite{Rey-Bellet2016}), cf.~Fig.~\ref{fig:Markov} (b) and (c). The minimum rate is then achieved for the dynamics that breaks local detailed balance, i.e., features at least one unidirectional transition (by allowing maximum values of the currents) and there always exist some observable for which the minimum is achieved only then. 
Those changes do not affect the value of the bound  Eq.~\eqref{eq:vOUR}, so that it can  saturate for such an observable  again only then.
Therefore, any dynamics that saturates Eq.~\eqref{eq:mOUR}  is not only away from equilibrium but breaks local detailed balance as it implies that Eq.~\eqref{eq:vOUR} saturates for all observables. \\

{ \emph{Importance for simulations}. The above considerations are crucial not only from the fundamental perspective but in the practical context of numerical simulations. 
	
	First,} for a given stationary distribution and fixed decay rates, the bound in Eq.~\eqref{eq:mOUR} is fully determined. Thus, with those constraints, any dynamics that saturates this bound is an optimal choice for stochastic simulations of the stationary distribution being estimated via occupation times, as it leads to the minimal fluctuation rates of those times.  When only a subset of observables is of interest, the saturation of Eq.~\eqref{eq:vOUR} is sufficient. We note, however, that other criteria are also employed, such as mixing times (see, e.g., Ref.~\cite{Kapfer2017}).

{ Second, in equilibrium simulations, high or low performance of the bound can be directly linked to coupling or decoupling between global and local timescales in the dynamics of system observables as follows. Under the similarity transformation with respect to $\sqrt{\mathbf{D}}$, Eq.~\eqref{eq:vOUR}  for the non-stationary component of the observable $\mathbf{o}'=\mathbf{Q}_\text{ss}\mathbf{o}$ becomes
	\begin{equation}\label{eq:vOUR_eq}
		2\sum_{k} \frac{\left(\widetilde{\bm \omega}_k^\text{T} \,\widetilde{\mathbf o}'\right)^2} {-w_k} \geq \sum_j  \frac{(\widetilde{\mathbf o}')_j^2}{ \lambda_j }.
	\end{equation}
	Here, $w_k$ are the eigenvalues of the symmetric matrix $\widetilde{\mathbf W}= \sqrt{\mathbf{D}}{\mathbf W} \sqrt{\mathbf{D}^{-1}}$ (which coincide with those of ${\mathbf W}$), $\widetilde{\bm \omega}_k$ the corresponding eigenvectors chosen normalised and real (these correspond both to $\widetilde{\bm \omega}_k^\text{L} =\sqrt{\mathbf{D}^{-1}} {\bm \omega}_k^\text{L}$ and $\widetilde{\bm \omega}_k^\text{R}=\sqrt{\mathbf{D}}{\bm \omega}_k^\text{R}$ for the left ${\bm \omega}_k^\text{L}$ and the right ${\bm \omega}_k^\text{R}$ eigenvectors of  ${\mathbf W}$~\footnote{We define left ${\bm \alpha}^\text{L}$ and right ${\bm \alpha}^\text{R}$ eigenvectors of a matrix ${\mathbf A}$ corresponding to an eigenvalue $a$ so that $ ({\bm \alpha}^\text{L})^\dagger {\mathbf A}=a({\bm \alpha}^\text{L})^\dagger$ and $ {\mathbf A} {\bm \alpha}^\text{R}=a {\bm \alpha}^\text{R}$ and the normalisation $({\bm \alpha}^\text{L})^\dagger {\bm \alpha}^\text{R}=1$ holds.}), and $\mathbf{\widetilde{o}'}=\sqrt{\mathbf{D}} \mathbf{o}'=\widetilde{\mathbf{Q}}_\text{ss}\widetilde{\mathbf{o}}$. Equation~\eqref{eq:vOUR_eq} is the inequality between: \emph{global timescales} given by $-1/w_k$ where $-w_k$ are the rates at which the dynamics eigenmodes decay exponentially in the average dynamics, and \emph{local timescales} given by  the lifetimes of the individual configurations $1/\lambda_j$, where $\lambda_j$ the decay rate of their survival probability in the stochastic dynamics. Those timescales are weighted with respect to how the observable decomposes in the corresponding: \emph{global basis} of the eigenmodes and \emph{local basis} of  the configurations, respectively (since only the non-stationary part is considered $k$ with $w_k=0$ does not contribute).  In particular, for $\widetilde{\mathbf o}'=\widetilde{\bm \omega}_k$ ($\mathbf{o}={\bm \omega}_k^R$), we obtain $-1/w_k\geq \sum_j (\widetilde{\bm \omega}_k)_j^2 /(2\lambda_j)$.
	These notions bear similarity to persistence and exchange times, for which the divergence of their ratio is considered a signature of decoupling of transport coefficients leading to breakdown of the Stokes-Einstein relation in supercooled liquids~\cite{Jung2004,Jung2005}.  We thus expect Eq.~\eqref{eq:vOUR_eq} to become a useful  diagnostic tool for emergent dynamic phenomena, in particular informing the choices of relevant system observables, cf.~Fig.~\ref{fig:Markov}(c). 
	Beyond equilibrium, the global timescales in the dynamics contribute to the rate of fluctuations for the time-integrated observable as $2\sum_{k} \text{Re}[(\widetilde{\mathbf o}')^\text{T} \widetilde{\bm \omega}_k^\text{R}
	(\widetilde{\bm \omega}_k^\text{L})^\dagger \,\widetilde{\mathbf o}'/(-w_k)]$, which replaces the left-hand side in Eq.~\eqref{eq:vOUR_eq} and gives rise to $\text{Re}(-1/w_k) \geq [\sum_j|(\widetilde{\bm \omega}_k^\text{R})_j|^2/(2\lambda_j)]/ [\sum_j|(\widetilde{\bm \omega}_k^\text{R})_j|^2]$ by summing the results  for $\widetilde{\mathbf o}'$ chosen  as $\text{Re}(\widetilde{\bm \omega}_k^R)$ and $\text{Im}(\widetilde{\bm \omega}_k^R)$ [${\mathbf o}$ chosen as  $\text{Re}({\bm \omega}_k^R)$ and $\text{Im}({\bm \omega}_k^R)$].
	%CHECK the positivity, the importance of the imaginary part, etc. !!!! Refer to the example, where we do no longer see a phase transition!!! 
} \\

%ABOUT that local unit of time, see what they say in references.... or is the dynamics discrete?
{ \emph{Connecting dynamic and static correlations}. Our results provide insights into dynamics timescales also with relevance to critical phenomena, where the divergence of static and dynamic correlations can be compared provided that an appropriate unit of time is chosen for continuous dynamics. For a given observable $\mathbf{o}$ such a unit could be  $1/\lambda_\mathbf{o}\equiv(\mathbf{Q}_\text{ss}\mathbf{o})^\text{T}\mathbf{B}^{-1}(\mathbf{Q}_\text{ss}\mathbf{o})/[(\mathbf{Q}_\text{ss}\mathbf{o})^\text{T}\mathbf{D}(\mathbf{Q}_\text{ss}\mathbf{o})]$ [i.e., the weighted average of configuration lifetimes $[\sum_j (\widetilde{\mathbf o}')_j^2/ \lambda_j]/[\sum_j  (\widetilde{\mathbf o}')_j^2]$, cf.~Eq.~\eqref{eq:vOUR_eq}].   Since $\mathbf{o}^\text{T}\mathbf{Q}_\text{ss}^\text{T}\mathbf{D} (\mathbf{Q}_\text{ss}\mathbf{o})$ coincides with the static variance  $\mathbf{o}^\text{T} \mathbf{\Delta}^2\mathbf{o}$ as $\mathbf{Q}_\text{ss}^\text{T}\mathbf{D} \mathbf{Q}_\text{ss}= (\mathbf{I}-\mathbf{p}_\text{ss}\mathbf{1}^\text{T})\mathbf{D} (\mathbf{I}-\mathbf{1}\mathbf{p}_\text{ss}^\text{T})=\mathbf{D}-\mathbf{p}_\text{ss}\mathbf{p}_\text{ss}^\text{T}=\bm{\Delta}^2$, 
	%	[or equivalently $\sum_j  (\widetilde{\mathbf o}')_j^2=\mathbf{o}^\text{T} \mathbf{\Delta}^2\mathbf{o}$]	
	Eq.~\eqref{eq:vOUR}  then yields
	\begin{equation} \label{eq:vOUR_var}
		\mathbf{o}^\text{T}	\mathbf{C}\mathbf{o}\geq \frac{1}{\lambda_\mathbf{o}} 	\mathbf{o}^\text{T} \mathbf{\Delta}^2\mathbf{o},
	\end{equation}
	so that the rate of dynamic variance of the time-integrated observable measured in the units of $1/\lambda_\mathbf{o}$ always exceeds the static variance for that observable, cf.~Fig.~\ref{fig:Markov}(c).
}
In general, the rates of the correlations between the occupation times can be related to the correlations in the stationary distribution via the maximal decay rate $\lambda_{\max}\equiv\max_{j}(\mathbf{\Lambda})_{jj}$ as
\begin{equation} \label{eq:mOUR_var}
	\mathbf{C}\geq \frac{1}{\lambda_{\max}} \mathbf{\Delta}^2.
\end{equation}
This follows from Eq.~\eqref{eq:mOUR}, since $\mathbf{B}^{-1}\geq \mathbf{D} /\lambda_{\max}$  { or from Eq.~\eqref{eq:vOUR_var} since $1/\lambda_\mathbf{o}\geq 1/\lambda_{\max}$. %and  $\mathbf{\Delta}^2=\mathbf{Q}_\text{ss}\mathbf{D} \mathbf{Q}_\text{ss}$.
	Eqs.~\eqref{eq:mOUR} and~\eqref{eq:vOUR_var}  reduce} to this bound when the decay rates are uniform, $(\mathbf{\Lambda})_{jj}=\lambda_{\max}$ for all $j$, but in general it is tighter, so that both its best and worst performance exceeds that of Eq.~\eqref{eq:mOUR_var}, cf.~Fig.~\ref{fig:Markov}(d).  
Bounds in terms of $L2$ norm for the observable shifted by its average can also be derived, see~\cite{SM}.

As we explain next, from { Eqs.~\eqref{eq:vOUR_var} and}~\eqref{eq:mOUR_var}, a \emph{dynamic exponent} for any observable can be defined to be non-negative,
\begin{equation} \label{eq:z}
	z_\mathbf{o}\geq 0,
\end{equation}
while from Refs.~\cite{Rey-Bellet2016,Kaiser2017}, this exponent is bounded from above by the corresponding exponent at equilibrium,  $z_\mathbf{o}^{\text{eq}}\geq z_\mathbf{o}$. Beyond fundamental importance to non-equilibrium statistical physics, these results provide a general framework to investigate breaking detailed balance for improved stochastic simulations (cf.~Refs.~\cite{Kapfer2017,Lei2019,Maggs2022}). 

Indeed, for the \emph{integrated autocorrelation time} $\tau_\mathbf{o}\equiv \mathbf{o}^\text{T}\mathbf{C} \mathbf{o}/ (\mathbf{o}^\text{T}\mathbf{\Delta}^2 \mathbf{o})/2$, which is the infinite time-integral of the observable autocorrelation %(which coincides with  $\mathbf{o}^\text{T}\mathbf{C} \mathbf{o}/ 2$)  
normalised by its variance in the stationary distribution~\cite{Sokal1997}, $z_\mathbf{o} \equiv \lim_{d\rightarrow\infty} \ln({\lambda_\mathbf{o}}\tau_\mathbf{o} )/\ln d$ captures its exponential divergence in the thermodynamic limit of ${d\rightarrow\infty}$~\footnote{We consider divergence with respect to $\ln d$ for generality. For systems such as $D$-dimensional lattices of volume $V=L^{D}$ with $n$ configurations per local site, so that $d=n^V$, dynamic exponents are typically calculated with respect to $V$ (cf.~Ref.~\cite{Sokal1997}). Then,  $z_\mathbf{o}$ needs to be multiplied by $\ln n$.}. Here,  we have chosen {$1/\lambda_\mathbf{o}$  as the unit of time, which can be replaced by $1/\lambda_{\max}$ to remove the need to calculate the former or its observable dependence; this can only increase the dynamical exponent}. We arrive at Eq.~\eqref{eq:z} as $\lambda_{\max}\tau_\mathbf{o}\geq \lambda_\mathbf{o}\tau_\mathbf{o}\geq 1/2$ from { Eqs.~\eqref{eq:vOUR_var} and~\eqref{eq:mOUR_var}.  
	
	Since Eq.~\eqref{eq:vOUR_var} coincides with  Eq.~\eqref{eq:vOUR}, the non-zero dynamical exponent $z_\mathbf{o}$, with $1/\lambda_\mathbf{o}$ as the unit of time, is the rate of the exponential decay of the vanishing performance of those bounds. Similarly, for $1/\lambda_\mathbf{o}$ chosen as the unit of time, $z_\mathbf{o}$ is the rate for the performance of Eq.~\eqref{eq:mOUR_var}  for a given system observable $\mathbf{o}$. 
	For equilibrium dynamics, the integrated autocorrelations time can be further recognised as a weighted average of the decay rates for the eigenmodes in the average dynamics [$[\sum_{k} (\widetilde{\bm \omega}_k^\text{T} \,\widetilde{\mathbf o}')^2/(-w_k)]/ [\sum_k(\widetilde{\bm \omega}_k^\text{T} \,\widetilde{\mathbf o}')^2]$, where $\sum_k(\widetilde{\bm \omega}_k^\text{T} \,\widetilde{\mathbf o}')^2=\mathbf{o}^\text{T} \mathbf{\Delta}^2\mathbf{o}$, cf.~Eq.~\eqref{eq:vOUR_eq}].  Then, $z_\mathbf{o}^{\text{eq}}$ is the rate in the exponential divergence of the ratio of this average of global timescales to the corresponding average of local timescales  $1/\lambda_{\mathbf{o}}$ or the shortest timescale $1/\lambda_{\max}$.}\\

\emph{Example I --- Saturation at finite size}. For the cycle dynamics with $(\mathbf{W})_{jk}= \lambda_j [-\delta_{jk}+x\delta_{(j-1)k}+(1-x)\delta_{(j+1)k}]$, cf.~Fig.~\ref{fig:Markov}(a), we have $\mathbf{p}_\text{ss}=\mathbf{\Lambda}^{-1}\mathbf{1}/\mathbf{1}^\text{T}\mathbf{\Lambda}^{-1}\mathbf{1}$, where $(\mathbf{\Lambda})_{jj}=\lambda_j$. Here, $0\leq x\leq 1$. Since $(\mathbf{p}_\text{ss})_j(\mathbf{W})_{j(j-1)}/x =(\mathbf{p}_\text{ss})_{j-1} (\mathbf{W})_{(j-1)j}/(1-x)$, the detailed balance is satisfied at $x=1/2$,  while $x=0$ and $x=1$ both lead to unidirectional transitions.

For uniform cycles, $\lambda_j=\lambda$, which ensures the translation symmetry.   Then, 
$\mathbf{p}_\text{ss}=\mathbf{1}/d$, while $\mathbf{W}$ and $\mathbf{C}$ are both diagonal in the basis of $(\bm{\omega}_k)_j=e^{  i  j k\frac{ 2 \pi }{d} }/\sqrt{d}$ with the corresponding eigenvalues $w_k=\lambda[-1+x e^{ i  k\frac{2 \pi }{d} }+(1-x)e^{-  i  k\frac{2 \pi}{d} }]=w_{d-k}^*$ and $c_k=2 [1-\cos(k\frac{2 \pi }{d})]/\{[1-\cos(k\frac{2 \pi }{d})]^2 +(1-2x)^2 \sin^2(k\frac{2 \pi }{d})\} /(d \lambda)=c_{d-k}$, respectively; $k=0,1,...,d-1$. On the other hand, $\mathbf{Q}_\text{ss}^\text{T}\mathbf{B}^{-1}\mathbf{Q}_\text{ss}=(\mathbf{I}-\mathbf{1}\mathbf{1}^\text{T}/d)/(d\lambda)$, so that it is diagonal with eigenvalues all equal $1/(d\lambda)$ (except for $0$ when $k=0$). 
{ Here, the rate of dynamic variance $\mathbf{o}^\text{T}	\mathbf{C}\mathbf{o}$ and the static variance $	\mathbf{o}^\text{T}	\bm{\Delta}^2\mathbf{o}$ are identical for $\mathbf{o}$ chosen as  $\text{Re}(\bm{\omega}_k)$ and $\text{Im}(\bm{\omega}_k)$ (since they coincide for $\bm{\omega}_k$ and $\bm{\omega}_{d-k}$), so for brevity we will use $\bm{\omega}_k$ to denote such observables.}
Therefore, the best and worst performance of the mOUR in~\eqref{eq:mOUR} is given by $(d \lambda) c_k$, with $k=1$ or $d-1$ in the best case, and $k=d/2$ for $d$ even and $(d\pm1)/2$ for $d$ odd in the worst case.
In particular, Eq.~\eqref{eq:mOUR} saturates for $x=0,1$, while Eq.~\eqref{eq:vOUR} can saturate for all $x$ when $d$  is even, cf.~Fig.~\ref{fig:Markov}(c).  The limiting {best and worst} performance when $d\rightarrow\infty$ is  $1$ and $(1-2x)^2$, so the bound remains faithful except at equilibrium. 

Interestingly, the best and worst performance of the asymptotic mOUR in Eq.~\eqref{eq:mOUR} appears universal for any decay rates, see Fig.~\ref{fig:Markov}(c). In particular, Eq.~\eqref{eq:mOUR} saturates for unidirectional cycles ($x=0,1$), which result is shown analytically for any $d$ in~\cite{SM}. Those dynamics can be viewed as optimal for stochastic simulations of stationary distributions  $\mathbf{p}_\text{{ss}}\propto \mathbf{\Lambda}^{-1}\mathbf{1}$. 

For uniform cycles,	   the bounds 	{ in Eqs.~\eqref{eq:vOUR} and~\eqref{eq:vOUR_var} as well as in Eqs.~\eqref{eq:mOUR} and~\eqref{eq:mOUR_var} coincide as $ \lambda_{\mathbf{o}}=\lambda_{\max}=\lambda$ and} $\mathbf{\Delta}^2 =(\mathbf{I}-\mathbf{1}\mathbf{1}^\text{T}/d)/d$,  cf.~Figs.~\ref{fig:Markov}(c) and~\ref{fig:Markov}(d). Since $\mathbf{p}_\text{ss}$ is analogous to the equilibrium distribution at  infinite temperature, the static criticality is absent in the thermodynamic limit. Nevertheless, dynamic criticality arises at equilibrium  ($x=1/2$).
The maximum exponent is $z^\text{eq}_{\max}=2$, as $\lambda\tau_{\bm{\omega}_k}$ diverges as $(d/k)^2/(2\pi^2)$ when $k/d\rightarrow 0$ or as $[d/(d-k)]^2/(2\pi^2)$ when $k/d\rightarrow 1$. The minimum exponent is $z^\text{eq}_{\min}=0$ since $\lambda\tau_{\bm{\omega}_k}\rightarrow 1/(1-\cos \phi)$ when $k/d\rightarrow \phi/(2\pi)$. The criticality is responsible for the vanishing performance of the mOUR for observables with $z_{\mathbf{o}}\neq0$, cf.~Fig.~\ref{fig:Markov}(c), and originates from the generator gap $(-w_1=-w_{d-1})$ closing { as the eigenvalues for which $k/d\rightarrow0$ or $1$ when $d\rightarrow \infty$ vanish as $-2\pi^2 k^2/d^2$ or $2\pi^2 (d-k)^2/d^2$}, while the eigenvalues of $\mathbf{\Delta}^2$ are all $1/d$ (except for $0$ corresponding to $\mathbf{1}$). If the performance of Eq.~\eqref{eq:mOUR} is universal for any decay rates also when $d\rightarrow\infty$,  the worst performance of both Eqs.~\eqref{eq:mOUR} and~\eqref{eq:mOUR_var} must reach $0$, cf.~Figs.~\ref{fig:Markov}(c) and~\ref{fig:Markov}(d), { signifying $z^\text{eq}_{\min}=2$ and $z^\text{eq}_{\min}\geq2$ for $1/\lambda_\mathbf{o}$ and $1/\lambda_{\max}$ as respective time units. 
	For uniform cycles, the real part of the gap close all the same away from equilibrium ($x\neq1/2$) when as $d\rightarrow\infty$ and so vanish the real parts of the eigenvalues with $k/d\rightarrow0$ or $1$, but their non-zero imaginary counterparts lead to 
	%$=\text{Im}(w_{1})
	%=\lambda (1-2x) \sin (2\pi/d) $, ensures that $\Re(-1/w_1)
	%=\text{Re}(-1/w_{d-1})
	%=[1-\cos(2\pi/d)]/ \{[1-\cos(2\pi/d)]^2+(1-2x)^2 \sin^2 (2\pi/d) \}/\lambda \rightarrow 1/(1-2x)^2/(2\lambda)
	%=\sin^2(\pi/d)/ [\sin^2(\pi/d)-x(1-x) \sin^2 (2\pi/d) ]/2/\lambda \rightarrow 1/(1-2x)^2/2$
	$\text{Re}(-1/w_k)=\sin^2(k\pi/d)/ [\sin^2(k\pi/d)-x(1-x) \sin^2 (2k\pi/d) ]/(2\lambda) \rightarrow 1/(1-2x)^2/(2\lambda)$
	and} so autocorrelations times remain finite as well, { $\lambda\tau_{\bm{\omega}_k}\rightarrow 1/(1-2x)^2/2$ when $k/d\rightarrow 0,1$ and} $\lambda\tau_{\bm{\omega}_k}\rightarrow(1-\cos \phi)/[(1-\cos \phi)^2 +(1-2x)^2\sin^2\phi]$ when $k/d\rightarrow \phi/(2\pi)$. {Thus, no signature of criticality arises, $z_{\mathbf{o}}=0$.  While the critical slowing down does not manifest itself in variance rates for system observables and the performance of the Eq.~\eqref{eq:vOUR_var},  the lower bound in Eq.~\eqref{eq:z} saturates in turn, so} those non-equilibrium dynamics could be viewed as optimal for stochastic simulations with respect to dynamic exponents.\\

\emph{Example II --- Saturation in thermodynamic limit}.  For $\mathbf{p}_\text{ss}$ such that $(\mathbf{p}_\text{ss})_{j}>0$ for all $j$, let $(\mathbf{W})_{jk}= \lambda [- \,\delta_{jk}+(\mathbf{p}_\text{ss})_{k}]$. Asymptotically, the equilibrium is reached, and the detailed balance holds,  $(\mathbf{p}_\text{ss})_{j}(\mathbf{W})_{jk}=\lambda(\mathbf{p}_\text{ss})_{j}(\mathbf{p}_\text{ss})_{k}=(\mathbf{p}_\text{ss})_{k}(\mathbf{W})_{kj}$ for $j\neq k$. Furthermore, as $\mathbf{W}=-\lambda \mathbf{Q}_\text{ss}$, we simply have  $\mathbf{R}=-(1/\lambda) \mathbf{Q}_\text{ss}$ and $\mathbf{C}=(2/\lambda) \mathbf{\Delta}^2 _\text{ss}$. 

Since $(\mathbf{\Lambda})_{jj}=\lambda[1- (\mathbf{p}_\text{ss})_{j}]$, we have $\lambda_{\max}=\lambda[1-\min_j (\mathbf{p}_\text{ss})_{j}]\leq \lambda$, as well as, $\lambda_{\max}\rightarrow \lambda$ when $d\rightarrow \infty$ [as $\min_j (\mathbf{p}_\text{ss})_{j}\leq 1/d \rightarrow 0$].  
Thus, the performance of both the bound in Eq.~\eqref{eq:mOUR_var} and the mOUR in Eq.~\eqref{eq:mOUR}, is no worse  than $1/2$. In the thermodynamic limit, for any observable, this value is exactly the performance of Eq.~\eqref{eq:mOUR_var}, which also coincides with that of Eq.~\eqref{eq:mOUR} provided that $\max_j (\mathbf{p}_\text{ss})_{j} \rightarrow 0$ (e.g., for uniform $\mathbf{p}_\text{ss}$).

For any $\mathbf{p}_\text{ss}$,  the dynamics above leads to $\lambda  \tau_{\mathbf{o}}=1$, so $z_{\mathbf{o}}=0$ and the bound in Eq.~\eqref{eq:z} saturates. As such, the outlined equilibrium dynamics could be viewed optimal. This can, however, be considered trivial since, e.g., in simulations where the next transition time is chosen according to the decay rate of a current configuration and the next configuration according to the transition rate, the latter rate is directly proportional to the stationary probability that is being sampled. 
This demonstrates further physical constraints are needed in the dynamics optimisation such as locality of transitions.\\

\emph{Relation to earlier work}. 
The uncertainty considered in Eq.~\eqref{eq:OUR} has a form analogous to those for functions of finite stochastic trajectories in the fluctuation-response inequality (FRI)~\cite{Dechant2020} 
and the universal TUR (uTUR)~\cite{Ziyin2023},  which are defined with respect to a reference process. 
In fact, the OURs can be obtained by considering the processes from the derivations in~\cite{SM} and then taking the long-time limit, as the FRI and uTUR reduce to the linear
FRI~\cite{Dechant2020}  in perturbative limits. 
The matrix bound in Eq.~\eqref{eq:mOUR}, however, does not follow from the multiparameter uTUR~\cite{Ziyin2023} (cf.~Ref.~\cite{Timpanaro2019}), which is a reference-dependent matrix bound of rank $1$,  and thus, even in the perturbative limit, necessarily not faithful for more than $2$ configurations.
Also,  the recently derived bounds for system observables on the ratio of their power spectrum to their variance in the stationary distribution~\cite{Dechant2023} do not reduce to Eq.~\eqref{eq:mOUR_var} in the limit of zero frequency.\\

\emph{Conclusions}. In this work, we introduced the uncertainty for time integrals of system observables as the ratio of their variance to the squared difference between their scaled average (or the average at final time) and the average in a reference distribution. 
We derived OURs as lower observable-independent bounds on such uncertainties and proceeded to show they stem from the mOUR, a lower reference-independent limit on occupation time correlation, which saturates for unidirectional cycles.
{	The high performance of the derived bounds can be associated with the optimality of given dynamics for stochastic simulations of its stationary distribution. The low or vanishing performance can be seen as a consequence of decoupling or divergence between global and local timescales, which is guaranteed to be witnessed at equilibrium thanks to how those timescales then affect fluctuations of system observables.}
It remains to be seen how those relations can be extended in an operational way to finite times (cf.~Refs.~\cite{Horowitz2017,Liu2020,Koyuk2020}) and time-dependent dynamics, e.g., for periodic driving (cf.~Ref.~\cite{Koyuk2019}), as well as when they can be formulated for Langevin dynamics (cf.~Ref.~\cite{Pietzonka2022}).\\

\emph{Acknowledgements}. K.~M. thanks R.~L.~Jack and J. P. Garrahan for helpful discussions.

%\bibliography{OUR_references.bib}
	

\onecolumngrid
\newpage
\clearpage

	\begin{center}
		\textbf{\large Supplemental Material: Occupation Uncertainty Relations}
	\end{center}
	%%%%%%%%%% Merge with supplemental materials %%%%%%%%%%
	%%%%%%%%%% Prefix a "S" to all equations, figures, tables and reset the counter %%%%%%%%%%
	\setcounter{equation}{0}
	\setcounter{figure}{0}
	\setcounter{table}{0}
	\setcounter{page}{1}
	\makeatletter
	\renewcommand{\theequation}{S\arabic{equation}}
	\renewcommand{\thefigure}{S\arabic{figure}}

\vspace{0.5cm}	
{In this Supplemental Material proofs of Eqs.~(1), (2), and (3) in the main text are given, together with the (formal) derivation of tighter bounds than those. The conditions for saturation of all the bounds are derived and shown to hold for unidirectional cycles. Finally, looser bounds in terms of L2 norm are presented.}

	\section{Derivation of Occupation Uncertainty Relations} \label{SM:OUR}
	
	Below we provide derivation of the Occupation Uncertainty Relation (OUR) in Eq.~(1) of the main text using the level-2.5 large  deviation results for Markov Processes~\cite{Maes2008,Bertini2015}. We then proceed to derive bounds tighter than OURs, which still, however, depend on the choice of a reference distribution.  
	
	\subsection{Proof of OURs}\label{SM:OUR_proof}
	
	According to the Large Deviation principle~\cite{denHollander2000,Touchette2009}, the time-integral $\overline{O}(t)/t$ converges to $\langle O\rangle_\text{ss}$ with the probability 1, while the probability of its taking a different value $ {o} $ asymptotically decays exponentially with time $t$ at the rate  $ I( {o} )$. Thanks to the contraction principle~\cite{denHollander2000,Touchette2009}, the rate function can be expressed as
	\begin{equation}\label{eq:l1}
		I( {o} )=\min_{\substack{\{\mu_j\}_j,\{n_{jk}\}_{jk:\,j\neq k}:\\\sum_{j}  \mu_j o_j= {o} }} I(\{\mu_{j}\}_{j},\{n_{jk}\}_{jk:\,j\neq k}),
	\end{equation} 
	where the so called level-2.5 rate function~\cite{Maes2008,Bertini2015} 
	\begin{equation}\label{eq:l2.5}
		I(\{\mu_{j}\}_{j},\{n_{jk}\}_{jk:\,j\neq k})\equiv \sum_{j}  \sum_{k\neq j} \left[n_{jk}\ln\frac{n_{jk}}{ \mu_j(\mathbf{W})_{jk}}-n_{jk}+\mu_j(\mathbf{W})_{jk}\right].
	\end{equation} 
	It represents the rate of the asymptotic decay of the probability of observing $\mu_j$ for the part of total time spent in $j$, i.e., the occupation time rate, and $n_{jk}$ for the total number of transitions from $j$ to $k$  divided by the total time, i.e., the transition number rate from $j$ to $k$. Thus, we consider $\mu_j\geq 0$ such that $\sum_{j} \mu_j=1$ and $n_{jk}\geq 0$ such that $\sum_{k\neq j} n_{jk}=\sum_{k\neq j} n_{kj}$ for all $j$. In particular, for the average values $\sum_j\mu_j=(\mathbf{p}_\text{ss})_j$ and $n_{jk}= (\mathbf{p}_\text{ss})_j(\mathbf{W})_{jk}$, we have  $ I(\{(\mathbf{p}_\text{ss})_j\}_{j},\{(\mathbf{p}_\text{ss})_j(\mathbf{W})_{jk}\}_{jk})=0$. For those values  $\sum_j\mu_j o_j=\langle O\rangle_\text{ss}$, so the minimum in Eq.~\eqref{eq:l1} for $ {o}  =\langle O\rangle_\text{ss}$ is no larger than $0$.  From the positivity of the rate function, we recover that $I(\langle O\rangle_\text{ss})=0$.  \\

	\emph{Derivation of Eq.~(1) in the main text}.  Let us consider the perturbative expansions around the average values for the rates of occupation times and transition numbers,
	\begin{equation}\label{eq:mu}
		\mu_j =(\mathbf{p}_\text{ss})_j+\epsilon (\boldsymbol{\delta}\mathbf{p})_j, \quad n_{jk}=(1+\epsilon\delta) (\mathbf{p}_\text{ss})_j(\mathbf{W})_{jk}.
	\end{equation}

	We assume that $\boldsymbol{\delta}\mathbf{p}^\text{T}\mathbf{1}=0$ so that  $\sum_{j} \mu_j=1$ as well as $\epsilon$ is small enough so that  $\mu_j\geq 0$ and $n_{jk}\geq 0$ [note that $n_{jk}$ are the average values uniformly rescaled by $(1+\epsilon\delta)$, so that the condition $\sum_{k\neq j} n_{jk}=\sum_{k\neq j} n_{kj}$ remains satisfied].
	
	The  level-2.5 rate function in Eq.~\eqref{eq:l2.5} expands as
	\begin{equation}\label{eq:l2.5OUR}
		I(\{\mu_{j}\}_{j},\{n_{jk}\}_{jk})
		=  \frac{\epsilon^2}{2}\sum_{j}\sum_{k\neq j} (\mathbf{p}_\text{ss})_j(\mathbf{W})_{jk} \left[ \frac{(\boldsymbol{\delta}\mathbf{p})_j}{(\mathbf{p}_\text{ss})_j}-\delta\right]^2 +\mathcal{O}(\epsilon^3)
		= \frac{\epsilon^2}{2} (\boldsymbol{\delta}\mathbf{p}-\delta\mathbf{p}_\text{ss})^\text{T}\mathbf{B}( \boldsymbol{\delta}\mathbf{p}-\delta\mathbf{p}_\text{ss})+\mathcal{O}(\epsilon^3),
	\end{equation}
	where we used the definition of $\mathbf{B}$ to introduce
	\begin{equation} \label{eq:B}
		(\mathbf{B})_{jj}=\sum_{k\neq j} \frac{(\mathbf{W})_{jk}}{(\mathbf{p}_\text{ss})_j},\quad (\mathbf{B})_{jk}=0\ \text{for}\ k\neq j,
	\end{equation}
	as defined in the main text. 
	In the first equality, we used  	$n_{jk}/[ \mu_j(\mathbf{W})_{jk}]=(1+\epsilon\delta)(\mathbf{p}_\text{ss})_j/[(\mathbf{p}_\text{ss})_j+\epsilon(\boldsymbol{\delta}\mathbf{p})_j] =1-\epsilon[ (\boldsymbol{\delta}\mathbf{p})_j/(\mathbf{p}_\text{ss})_j-\delta] +\epsilon^2[ - (\boldsymbol{\delta}\mathbf{p})_j/(\mathbf{p}_\text{ss})_j -\delta] (\boldsymbol{\delta}\mathbf{p})_j/(\mathbf{p}_\text{ss})_j+ \mathcal{O}(\epsilon^3)$, so that $\ln \{ n_{jk}/[ \mu_j(\mathbf{W})_{jk}] \}=-\epsilon[(\boldsymbol{\delta}\mathbf{p})_j/(\mathbf{p}_\text{ss})_j-\delta] +\epsilon^2[ (\boldsymbol{\delta}\mathbf{p})_j/(\mathbf{p}_\text{ss})_j-\delta ](\boldsymbol{\delta}\mathbf{p})_j/(\mathbf{p}_\text{ss})_j-\epsilon^2[ (\boldsymbol{\delta}\mathbf{p})_j/(\mathbf{p}_\text{ss})_j-\delta]^2/2+\mathcal{O}(\epsilon^3)$, while $-n_{jk}+\mu_j(\mathbf{W})_{jk}=\epsilon (\mathbf{p}_\text{ss})_j(\mathbf{W})_{jk}[ (\boldsymbol{\delta}\mathbf{p})_j/(\mathbf{p}_\text{ss})_j-\delta]$.
	
	Meanwhile, the corresponding rate of $\overline{O}(t)/t$ is
	\begin{equation}\label{eq:o0}
		{o} =\sum_j\mu_j o_j=\langle O\rangle_\text{ss}+\epsilon \boldsymbol{\delta}\mathbf{p}^\text{T}\mathbf{o}.
	\end{equation}
	Thus, the rate function in Eq.~\eqref{eq:l1} expands as
	\begin{equation}\label{eq:l1OUR}
		I( {o} )=  \frac{1}{2} I''(\langle O\rangle_\text{ss}) ( {o} -\langle O\rangle_\text{ss})^2+\mathcal{O}(\epsilon^2)=
		\frac{\epsilon^2}{2}\frac{(\boldsymbol{\delta}\mathbf{p}^\text{T}\mathbf{o})^2}{\mathbf{o}^\text{T}\mathbf{C}\mathbf{o}}+\mathcal{O}(\epsilon^3),	
	\end{equation}
	where in the second equality we used fact that the second derivative of the rate function at  $ {o} =\langle O\rangle_\text{ss}$ equals the inverse of the variance rate for $\overline{O}(t)$~\cite{denHollander2000,Touchette2009}, which corresponds to the central limit theorem for $\sqrt{t}[\overline{O}(t)/t-  \langle O\rangle_\text{ss}]$.  
	
	Due to Eq.~\eqref{eq:o0}, the rate function in Eq.~\eqref{eq:l1OUR} is bounded from above by the rate function in Eq.~\eqref{eq:l2.5OUR}, so that 
	\begin{equation} \label{eq:OUR_delta_full}
		I( {o} )=\frac{\epsilon^2}{2}\frac{(\boldsymbol{\delta}\mathbf{p}^\text{T}\mathbf{o})^2}{\mathbf{o}^\text{T}\mathbf{C}\mathbf{o}}+\mathcal{O}(\epsilon^3)\leq \frac{\epsilon^2}{2} (\delta\mathbf{p}_\text{ss}-\boldsymbol{\delta}\mathbf{p})^\text{T}\mathbf{B}( \delta \mathbf{p}_\text{ss}-\boldsymbol{\delta}\mathbf{p})+\mathcal{O}(\epsilon^3)
	\end{equation}	
	and considering the leading second order, we obtain  
	\begin{equation} \label{eq:OUR_delta}
		\frac{(\boldsymbol{\delta}\mathbf{p}^\text{T}\mathbf{o})^2}{\mathbf{o}^\text{T}\mathbf{C}\mathbf{o}} \leq  (\boldsymbol{\delta}\mathbf{p}-\delta\mathbf{p}_\text{ss})^\text{T}\mathbf{B}( \boldsymbol{\delta}\mathbf{p}-\delta\mathbf{p}_\text{ss}) .
	\end{equation}	
	Considering the choice of $\delta$ leading to  the minimum value of the right-hand side, we further obtain 
	\begin{equation} \label{eq:OUR_delta_opt_full}
		I( {o} )=\frac{\epsilon^2}{2}\frac{(\boldsymbol{\delta}\mathbf{p}^\text{T}\mathbf{o})^2}{\mathbf{o}^\text{T}\mathbf{C}\mathbf{o}}+\mathcal{O}(\epsilon^3)\leq \frac{\epsilon^2}{2} \boldsymbol{\delta}\mathbf{p}^\text{T}\!\left(\mathbf{B}- \frac{\mathbf{B}\mathbf{p}_\text{ss}\mathbf{p}_\text{ss}^\text{T}\mathbf{B} }{\mathbf{p}_\text{ss}^\text{T}\mathbf{B}\mathbf{p}_\text{ss}}\right)\!\boldsymbol{\delta}\mathbf{p}+\mathcal{O}(\epsilon^3)
	\end{equation}	
	and
	\begin{equation} \label{eq:OUR_delta_opt}
		\frac{(\boldsymbol{\delta}\mathbf{p}^\text{T}\mathbf{o})^2}{\mathbf{o}^\text{T}\mathbf{C}\mathbf{o}}\leq \boldsymbol{\delta}\mathbf{p}^\text{T}\!\left(\mathbf{B}- \frac{\mathbf{B}\mathbf{p}_\text{ss}\mathbf{p}_\text{ss}^\text{T}\mathbf{B} }{\mathbf{p}_\text{ss}^\text{T}\mathbf{B}\mathbf{p}_\text{ss}}\right)\!\boldsymbol{\delta}\mathbf{p}.
	\end{equation}	
	The  minimum is achieved for  
	\begin{equation}\label{eq:delta}
		\delta = \frac{\mathbf{p}_\text{ss}^\text{T}\mathbf{B}\boldsymbol{\delta}\mathbf{p}} {\mathbf{p}_\text{ss}^\text{T}\mathbf{B}\mathbf{p}_\text{ss}}. 
	\end{equation}
	Eq.~(1) in the main text is finally obtained by inverting both sides of the inequality and replacing $\boldsymbol{\delta}\mathbf{p}=\mathbf{p}_\text{ss}-\mathbf{p}$.

	\subsection{Tighter bounds}\label{SM:OUR_other}
	
	We now derive other bounds on the uncertainty of system observables. Namely, 
	\begin{equation}\label{eq:OUR2}
		\lim_{t\,\rightarrow \infty} \,\frac{1}{t}\,\frac{\Delta^2 \overline{O}(t)}{[\langle \overline{O}(t)\rangle/t -\langle O\rangle]^2}
		=\frac{\mathbf{o}^\text{T} \mathbf{C}  \mathbf{o}}{[(\mathbf{p}_\text{ss}-\mathbf{p})^\text{T}\mathbf{o} ]^2}
		\geq \frac{1} {(\mathbf{p}_\text{ss}-\mathbf{p})^\text{T} (\mathbf{B}-\boldsymbol{\delta}\mathbf{B}) (\mathbf{p}_\text{ss}-\mathbf{p})-\frac{[(\mathbf{p}_\text{ss}-\mathbf{p})^\text{T}(\mathbf{B}-\boldsymbol{\delta}\mathbf{B}){\mathbf{p}_\text{ss}}]^2}{\mathbf{p}_\text{ss}^\text{T}(\mathbf{B}-\boldsymbol{\delta}\mathbf{B}){\mathbf{p}_\text{ss}}}},
	\end{equation}
	where [cf.~Eq.~\eqref{eq:B}]
	\begin{equation} \label{eq:deltaB}
		(\boldsymbol{\delta}\mathbf{ B})_{jj}\equiv \sum_{k\neq j}\frac{(\mathbf{p}_\text{ss})_k}{(\mathbf{p}_\text{ss})_j} \frac{(\mathbf{W})_{jk}(\mathbf{W})_{kj} }{(\mathbf{p}_\text{ss})_j(\mathbf{W})_{jk}+(\mathbf{p}_\text{ss})_k(\mathbf{W})_{kj}},\quad (\boldsymbol{\delta}\mathbf{ B})_{jk}\equiv  \frac{(\mathbf{W})_{jk}(\mathbf{W})_{kj} }{(\mathbf{p}_\text{ss})_j(\mathbf{W})_{jk}+(\mathbf{p}_\text{ss})_k(\mathbf{W})_{kj}}\ \text{for}\ j\neq k.
	\end{equation}
	These bounds still depend on the choice of the reference distribution $\mathbf{p}$, as is the case for OURs in Eq.~(1) of the main text. They are tighter than OURs, as $\boldsymbol{\delta}\mathbf{ B}$ is positive definite. This can be seen as a consequence of these bounds following from the level-2.5 rate function for Markov processes for occupations and currents only~\cite{Bertini2015a} which, by the contraction principle~\cite{denHollander2000,Touchette2009}, takes values below the level-2.5 rate function for occupations and fluxes exploited in Sec.~\ref{SM:OUR_proof}.\\

	\emph{Derivation of Eq.~\eqref{eq:OUR2}}. Let us consider the following perturbative expansions around the average value for the rates of occupation times transition numbers,
	\begin{equation}\label{eq:mu2}
		\mu_j =(\mathbf{p}_\text{ss})_j+\epsilon (\boldsymbol{\delta}\mathbf{p})_j, \quad n_{jk}= (1+\epsilon \delta)(\mathbf{p}_\text{ss})_j(\mathbf{W})_{jk} + \epsilon\delta_{jk}.
	\end{equation}
	Here, we assume that $\epsilon$ is small enough so that  $\mu_j\geq 0$ and $n_{jk}\geq 0$, while $\boldsymbol{\delta}\mathbf{p}^\text{T}\mathbf{1}=0$ so that  $\sum_{j} \mu_j=1$, as well as $\delta_{jk}=\delta_{kj}$, so that  $\sum_{k\neq j} n_{jk}=\sum_{k\neq j} n_{kj}$ holds.
	
	The rate function in Eq.~\eqref{eq:l2.5} then expands as [cf.~Eq.~\eqref{eq:l2.5OUR}]
	\begin{equation}\label{eq:l2.5OUR2}
		I(\{\mu_{j}\}_{j},\{n_{jk}\}_{jk:\,j\neq k})
		=  \frac{\epsilon^2}{2}\sum_{j}\sum_{k\neq j} (\mathbf{p}_\text{ss})_j(\mathbf{W})_{jk} \left[ \frac{(\boldsymbol{\delta}\mathbf{p})_j}{(\mathbf{p}_\text{ss})_j}-\delta-\frac{\delta_{jk} }{(\mathbf{p}_\text{ss})_j(\mathbf{W})_{jk}} \right]^2 +\mathcal{O}(\epsilon^3).
	\end{equation}
	This bounds from above the expansion of the rate function [as in Eq.~\eqref{eq:l1OUR}] for the corresponding time-integral rate $ {o} $ [as in Eqs.~\eqref{eq:o0}], so that in the leading second order we obtain 
	\begin{equation} \label{eq:OUR2_deltas}
		\frac{(\boldsymbol{\delta}\mathbf{p}^\text{T}\mathbf{o})^2}{\mathbf{o}^\text{T}\mathbf{C}\mathbf{o}} \leq \sum_{j}\sum_{k\neq j} (\mathbf{p}_\text{ss})_j(\mathbf{W})_{jk} \left[ \frac{(\boldsymbol{\delta}\mathbf{p})_j}{(\mathbf{p}_\text{ss})_j}-\delta-\frac{\delta_{jk} }{(\mathbf{p}_\text{ss})_j(\mathbf{W})_{jk}} \right]^2 .
	\end{equation}

	Considering the choice of $\{\delta_{jk}\}_{jk:\,j<k}$ leading to the minimum value of the right-hand side in Eq.~\eqref{eq:OUR2_deltas}, we get [cf.~\eqref{eq:OUR2_delta}]
	\begin{equation} \label{eq:OUR2_delta0}
		\frac{(\boldsymbol{\delta}\mathbf{p}^\text{T}\mathbf{o})^2}{\mathbf{o}^\text{T}\mathbf{C}\mathbf{o}} \leq \sum_{j}\sum_{k\neq j} (\mathbf{p}_\text{ss})_j(\mathbf{W})_{jk} \left[ \frac{(\boldsymbol{\delta}\mathbf{p})_j}{(\mathbf{p}_\text{ss})_j}-\delta \right]^2 - \frac{1}{2}	\sum_{j}\sum_{k\neq  j}\frac{(\mathbf{p}_\text{ss})_j(\mathbf{W})_{jk}(\mathbf{p}_\text{ss})_k(\mathbf{W})_{kj} }{(\mathbf{p}_\text{ss})_j(\mathbf{W})_{jk}+(\mathbf{p}_\text{ss})_k(\mathbf{W})_{kj}} \left[ \frac{(\boldsymbol{\delta}\mathbf{p})_j}{(\mathbf{p}_\text{ss})_j}+\frac{(\boldsymbol{\delta}\mathbf{p})_k}{(\mathbf{p}_\text{ss})_k}-2\delta \right]^2 ,
	\end{equation}	
	which is achieved for
	\begin{equation}\label{eq:delta_jk}
		\delta_{jk}=\left[\frac{(\boldsymbol{\delta}\mathbf{p})_j}{(\mathbf{p}_\text{ss})_j}+\frac{(\boldsymbol{\delta}\mathbf{p})_k}{(\mathbf{p}_\text{ss})_k} -2\delta\right] \frac{(\mathbf{p}_\text{ss})_j(\mathbf{W})_{jk}(\mathbf{p}_\text{ss})_k(\mathbf{W})_{kj}}{ (\mathbf{p}_\text{ss})_j(\mathbf{W})_{jk}+(\mathbf{p}_\text{ss})_k(\mathbf{W})_{kj}}.
	\end{equation} 
	This result can be rewritten in a matrix form using the definitions in Eqs.~\eqref{eq:B} and~\eqref{eq:deltaB}, as
	\begin{equation} \label{eq:OUR2_delta}
		\frac{(\boldsymbol{\delta}\mathbf{p}^\text{T}\mathbf{o})^2}{\mathbf{o}^\text{T}\mathbf{C}\mathbf{o}} \leq 	(\boldsymbol{\delta}\mathbf{p}-\delta\mathbf{p}_\text{ss})^\text{T}\left(\mathbf{B}-\boldsymbol{\delta}\mathbf{ B}\right)( \boldsymbol{\delta}\mathbf{p}-\delta\mathbf{p}_\text{ss}).
	\end{equation}	
	
	Finally,
	considering the choice of $\delta$ leading to the minimum value of the right-hand side in Eq.~\eqref{eq:OUR2_delta}, we obtain [cf.~Eq.~\eqref{eq:OUR_delta_opt}]
	\begin{equation} \label{eq:OUR2_delta_opt}
		\frac{(\boldsymbol{\delta}\mathbf{p}^\text{T}\mathbf{o})^2}{\mathbf{o}^\text{T}\mathbf{C}\mathbf{o}} \leq \boldsymbol{\delta}\mathbf{p}^\text{T}\!\left[\mathbf{B}-\boldsymbol{\delta}\mathbf{B}- \frac{(\mathbf{B}-\boldsymbol{\delta}\mathbf{B})\mathbf{p}_\text{ss}\mathbf{p}_\text{ss}^\text{T}(\mathbf{B}-\boldsymbol{\delta}\mathbf{B})}{\mathbf{p}_\text{ss}^\text{T}(\mathbf{B}-\boldsymbol{\delta}\mathbf{B})\mathbf{p}_\text{ss}}\right]\!\boldsymbol{\delta}\mathbf{p} .
	\end{equation}	
	The  minimum is achieved for 
	\begin{equation}\label{eq:delta2} \delta=\frac{\mathbf{p}_\text{ss}^\text{T}(\mathbf{B}-\boldsymbol{\delta}\mathbf{B})\boldsymbol{\delta}\mathbf{p}}{ \mathbf{p}_\text{ss}^\text{T}(\mathbf{B}-\boldsymbol{\delta}\mathbf{B})\mathbf{p}_\text{ss}}. 
	\end{equation}
	Eq.~\eqref{eq:OUR2} above is obtained by inverting both sides of this inequality and replacing $\boldsymbol{\delta}\mathbf{p}=\mathbf{p}_\text{ss}-\mathbf{p}$. 
	\\
	
	\emph{Comments}. 
	We note that the expansions  for the rates of occupation times and transitions numbers in Eq.~\eqref{eq:mu2} include those used in the derivation of OURs in Sec.~\ref{SM:OUR_proof} as a special case where $\delta_{jk}=0$, cf.~Eq.~\eqref{eq:mu}. Consequently, the bounds in Eq.~\eqref{eq:OUR2} are as good or better than the OURs. This directly manifest in $\boldsymbol{\delta}\mathbf{B}$ being a positive semi-definite  matrix [cf.~Eqs.~\eqref{eq:OUR2_delta0} and~\eqref{eq:OUR2_delta}]. 
	
	Alternatively, the improved quality can be seen from the relation between different level-2.5 rate functions for Markov processes (cf.~Ref.~\cite{Barato2015b}): for occupations and fluxes~\cite{Maes2008,Bertini2015} and for occupations and currents only~\cite{Bertini2015a}. Namely, the expansions for the rates of  transitions numbers in Eq.~\eqref{eq:mu2} can  be expressed as
	\begin{equation}
		n_{jk}+n_{kj}= {(\mathbf{p}_\text{ss})_j(\mathbf{W})_{jk}+(\mathbf{p}_\text{ss})_k(\mathbf{W})_{kj}}+\epsilon\tilde{\delta}_{jk},\quad n_{jk}-n_{kj}=(1+\epsilon\delta ) \left[(\mathbf{p}_\text{ss})_j(\mathbf{W})_{jk}-(\mathbf{p}_\text{ss})_k(\mathbf{W})_{kj}\right],
	\end{equation}
	where $\tilde{\delta}_{jk}=\delta_{jk}+\delta [(\mathbf{p}_\text{ss})_j(\mathbf{W})_{jk}+(\mathbf{p}_\text{ss})_k(\mathbf{W})_{kj}] $. 
	Since both $\delta_{jk}$ and  $\delta$ are optimised in the derivation of Eq.~\eqref{eq:OUR2}, this is equivalent to optimising $\tilde{\delta}_{jk}$ and $\delta$. First, optimisating $\tilde{\delta}_{jk}$ corresponds to choosing optimal symmetric fluxes $n_{jk}+n_{kj}$ leading, by the contraction principle, to the rate function for occupation time rates $\mu_j$ and asymmetric fluxes (currents) $n_{jk}-n_{kj}$. Second, the remaining fluxes are uniformly rescaled from their average values, while in the derivation of OURs all fluxes were rescaled uniformly. Finally, from this procedure it also follows that the bounds in Eq.~\eqref{eq:OUR2} can saturate for any cycles (for example, as in Fig. 1 of the main text) and equilibrium dynamics without asymptotic currents. See Sec.~\ref{SM:sat2}.

	\section{Derivation of matrix Occupation Uncertainty Relation} \label{SM:mOUR}
	
	We now derive Eq.~(2), or equivalently, the matrix Occupation Uncertainty Relation (mOUR) in Eq.~(3),  both in the main text. We also comment on steps required to obtain matrix bounds from tighter relations derived in Sec.~\ref{SM:OUR_other} of this Supplemental Material.
	
	\subsection{Proof of mOUR} \label{SM:mOUR_proof}
	
	Here, we prove Eq.~(2) of the main text. We also explain on how Eq.~(1) can be recovered from that result.\\
	
	\emph{Derivation of Eq.~(2) in the main text}. From the derivation of Eq.~(1) of the main text in Sec.~\ref{SM:OUR_proof}, we have [cf.~Eq.~\eqref{eq:OUR_delta}]
	\begin{equation} \label{eq:OUR_delta_inv}
		\mathbf{o}^\text{T}\mathbf{C}\mathbf{o}\geq \frac{(\boldsymbol{\delta}\mathbf{p}^\text{T}\mathbf{o})^2}{(\boldsymbol{\delta}\mathbf{p}-\delta\mathbf{p}_\text{ss})^\text{T}\mathbf{B}(\boldsymbol{\delta}\mathbf{p}-\delta \mathbf{p}_\text{ss} )},
	\end{equation} 
	where  $\boldsymbol{\delta}\mathbf{p}^\text{T}\mathbf{1}=0$. Then, for
	\begin{equation}\boldsymbol{\delta}\tilde{\mathbf{p}} \equiv \boldsymbol{\delta}\mathbf{p} -\delta\mathbf{p}_\text{ss},
	\end{equation}
	we have $\boldsymbol{\delta}\tilde{\mathbf{p}}^\text{T} \mathbf{Q}_\text{ss}=\boldsymbol{\delta}\mathbf{p}^\text{T}$,
	so we can rewrite Eq.~\eqref{eq:OUR_delta_inv} as 
	\begin{equation}\label{eq:OUR_delta_inv_tilde}
		\mathbf{o}^\text{T}\mathbf{C}\mathbf{o}\geq\frac{(\boldsymbol{\delta}{\tilde{\mathbf{p}}}^\text{T}\mathbf{Q}_\text{ss}\mathbf{o})^2}{\boldsymbol{\delta}{\tilde{\mathbf{p}}}^\text{T}\mathbf{B}\boldsymbol{\delta}\tilde{\mathbf{p}} }.
	\end{equation}
	Maximising the right-hand side with respect to $\boldsymbol{\delta}\tilde{\mathbf{p}}$, we obtain
	\begin{equation}\label{eq:vOUR_proof}
		\mathbf{o}^\text{T}\mathbf{C}\mathbf{o}\geq\mathbf{o}^\text{T}\mathbf{Q}_\text{ss}^\text{T}  \mathbf{B}^{-1} \mathbf{Q}_\text{ss}\mathbf{o},
	\end{equation}
	which directly coincides with Eq.~(2) of the main text.
	Equation~\eqref{eq:vOUR_proof} can be obtained by considering the weighted scalar product $\langle \mathbf{u},\mathbf{v}\rangle\equiv\mathbf{u}^\text{T} \mathbf{B} \mathbf{v}$, which is well defined since $\mathbf{B}>\mathbf{0}$. Indeed, the right-hand side of  Eq.~\eqref{eq:OUR_delta_inv_tilde} can be expressed as $(\langle \boldsymbol{\delta}\tilde{\mathbf{p}}, \mathbf{B}^{-1} \mathbf{Q}_\text{ss} \mathbf{o}\rangle)^2/ \langle \boldsymbol{\delta}\tilde{\mathbf{p}},\boldsymbol{\delta}\tilde{\mathbf{p}}\rangle $. This is bounded from above by $\langle \mathbf{B}^{-1} \mathbf{Q}_\text{ss} \mathbf{o}, \mathbf{B}^{-1} \mathbf{Q}_\text{ss} \mathbf{o} \rangle=\mathbf{o}^\text{T}\mathbf{Q}_\text{ss}^\text{T}  \mathbf{B}^{-1} \mathbf{Q}_\text{ss}\mathbf{o}$  via Cauchy-Schwarz inequality that saturates when the scalar product arguments are parallel, that is, for
	\begin{equation} \label{eq:deltaptilde}
		\boldsymbol{\delta}\tilde{\mathbf{p}}\propto\mathbf{B}^{-1} \mathbf{Q}_\text{ss} \mathbf{o}.
	\end{equation}
	We note that this optimisation is valid, as there is no restriction on  $\boldsymbol{\delta}\tilde{\mathbf{p}}$ (in contrast to $\boldsymbol{\delta}\mathbf{p}$ with ${\boldsymbol{\delta}\mathbf{p}}^\text{T}\mathbf{1}=0$). It corresponds to optimising both $\boldsymbol{\delta}\mathbf{p}$ and $\delta$ in Eq.~\eqref{eq:OUR_delta_inv} (as ${\boldsymbol{\delta}\tilde{\mathbf{p}}}^\text{T}\mathbf{1}=-\delta$), which results in 
	\begin{equation}\label{eq:deltap}
		\boldsymbol{\delta}\mathbf{p}=\mathbf{Q}_\text{ss}^\text{T}\boldsymbol{\delta}\tilde{\mathbf{p}}\propto\mathbf{B}^{-1} \mathbf{Q}_\text{ss} \mathbf{o} - (\mathbf{1}^\text{T}\mathbf{B}^{-1} \mathbf{Q}_\text{ss} \mathbf{o})\mathbf{p}_\text{ss} 
	\end{equation} 
	and $\delta =-{\boldsymbol{\delta}\tilde{\mathbf{p}}}^\text{T}\mathbf{1} \propto \mathbf{1}^\text{T}\mathbf{B}^{-1} \mathbf{Q}_\text{ss} \mathbf{o}$ with the same proportionality constant as $\boldsymbol{\delta}\mathbf{p}$ [this coincides with the value of $\delta$ in Eq.~\eqref{eq:delta}  used for Eq.~\eqref{eq:OUR_delta_opt}].
	
	Finally, we note that  Eq.~\eqref{eq:vOUR_proof} also implies a perturbative bound for the level-1 rate function in Eq.~\eqref{eq:l1} [cf.~Eq.~\eqref{eq:OUR_delta_full}]
	\begin{equation} \label{eq:mOUR_full}
		I( {o} )\leq\frac{1}{2}\frac{({o}-\langle O\rangle_\text{ss})^2}{\mathbf{o}^\text{T}\mathbf{Q}_\text{ss}^\text{T}  \mathbf{B}^{-1} \mathbf{Q}_\text{ss}\mathbf{o}}+\mathcal{O}[({o}-\langle O\rangle_\text{ss})^3].
	\end{equation}	
	\\
	
	\emph{Connections between OURs and mOUR}. From the derivation above, it follows that in the main text Eq.~(2) can be obtained from Eq.~(1) by considering a reference distribution $\mathbf{p}=\mathbf{p}_\text{ss}+\boldsymbol{\delta}\mathbf{p}$, where $\boldsymbol{\delta}\mathbf{p}$ is given by Eq.~\eqref{eq:deltap}, so that
	\begin{equation}\label{eq:p}
		\mathbf{p}=\mathbf{p}_\text{ss}+\epsilon\left[\mathbf{B}^{-1} \mathbf{Q}_\text{ss} \mathbf{o} - (\mathbf{1}^\text{T}\mathbf{B}^{-1} \mathbf{Q}_\text{ss} \mathbf{o})\mathbf{p}_\text{ss} \right].
	\end{equation}
	This corresponds to a reference $\mathbf{p}$ leading to the highest quality of the OUR for a given observable $\mathbf{o}$.
	
	Conversely, Eq.~(1) can be recovered from Eq.~(2) in the main text, by observing that from the latter [or Eq.~\eqref{eq:vOUR_proof}] we obtain the following bound on the uncertainty for an observable $\mathbf{o}$ and reference $\mathbf{p}$, 
	\begin{equation} \label{eq:mOUR_OUR}
		\frac{\mathbf{o}^\text{T}\mathbf{C}\mathbf{o}  } {[(\mathbf{p}-\mathbf{p}_\text{ss})^\text{T}\mathbf{o}]^2} \geq \frac{\mathbf{o}^\text{T}\mathbf{Q}_\text{ss}^\text{T}\mathbf{B}^{-1} \mathbf{Q}_\text{ss}\mathbf{o}} { [(\mathbf{p}-\mathbf{p}_\text{ss})^\text{T}\mathbf{o})]^2}.
	\end{equation}
	Denoting $\boldsymbol{\delta}\mathbf{p}=\mathbf{p}-\mathbf{p}_\text{ss}$ (so that $\boldsymbol{\delta}\mathbf{p}^\text{T}\mathbf{1}=0$ and thus $\boldsymbol{\delta}\mathbf{p}^\text{T}\mathbf{Q}_\text{ss}=\boldsymbol{\delta}\mathbf{p}^\text{T}$), we further have
	\begin{equation} \label{eq:mOUR_OUR_delta}
		\frac{\mathbf{o}^\text{T}\mathbf{C}\mathbf{o}  } {(\boldsymbol{\delta}\mathbf{p}^\text{T}\mathbf{o})^2} \geq\frac{(\mathbf{Q}_\text{ss}\mathbf{o})^\text{T}\mathbf{B}^{-1} (\mathbf{Q}_\text{ss} \mathbf{o})} { [
			\boldsymbol{\delta}\mathbf{p}^\text{T}(\mathbf{Q}_\text{ss}\mathbf{o})]^2}= \frac{(\mathbf{Q}_\text{ss}\mathbf{o})^\text{T}\mathbf{B}^{-1} (\mathbf{Q}_\text{ss} \mathbf{o})} { [(\boldsymbol{\delta}\mathbf{p}-\delta \mathbf{p}_\text{ss})^\text{T}(\mathbf{Q}_\text{ss}\mathbf{o})]^2}\geq \frac{1}{(\boldsymbol{\delta}\mathbf{p}-\delta \mathbf{p}_\text{ss})^\text{T}\mathbf{B}(\boldsymbol{\delta}\mathbf{p}-\delta \mathbf{p}_\text{ss})}.
	\end{equation} 
	Here, in the equality we used   $\mathbf{Q}_\text{ss}^2=\mathbf{Q}_\text{ss}$ together with $\mathbf{p}_\text{ss}^\text{T}  \mathbf{Q}_\text{ss}=\mathbf{0}$. The second inequality follows from Cauchy-Schwarz inequality for the scalar product $\langle \mathbf{u},\mathbf{v}\rangle\equiv\mathbf{u}^\text{T} \mathbf{B}^{-1} \mathbf{v}$. This inequality can be  saturated provided that $\mathbf{Q}_\text{ss}\mathbf{o}\propto \mathbf{B}(\boldsymbol{\delta}\mathbf{p}-\delta \mathbf{p}_\text{ss})^\text{T}$. As $\mathbf{Q}_\text{ss}\mathbf{o}=\mathbf{Q}_\text{ss}^2\mathbf{o}$, this is valid when  $\mathbf{B}(\boldsymbol{\delta}\mathbf{p}-\delta \mathbf{p}_\text{ss})=\mathbf{Q}_\text{ss} \mathbf{B}(\boldsymbol{\delta}\mathbf{p}-\delta \mathbf{p}_\text{ss})
	=\mathbf{B}(\boldsymbol{\delta}\mathbf{p}-\delta \mathbf{p}_\text{ss})-\mathbf{1}\mathbf{p}_\text{ss}^\text{T}\mathbf{B}(\boldsymbol{\delta}\mathbf{p}-\delta \mathbf{p}_\text{ss})$ which requires  $\delta =\mathbf{p}_\text{ss}^\text{T}\mathbf{B}\boldsymbol{\delta}\mathbf{p}/ (\mathbf{p}_\text{ss}^\text{T}\mathbf{B}\mathbf{p}_\text{ss})$.  
	Note that Eq.~\eqref{eq:mOUR_OUR_delta} coincides with Eq.~\eqref{eq:OUR_delta}, which for such $\delta$ leads to Eq.~(1) of the main text [cf.~Eq.~\eqref{eq:OUR_delta_opt}], so the OUR is recovered.  
	
	Finally, the OUR for the reference $\mathbf{p}=\boldsymbol{\delta}\mathbf{p}+\mathbf{p}_\text{ss}$ is obtained from Eq.~\eqref{eq:mOUR_OUR} by considering the value attained by the bound for an observable $\mathbf{o}$ such that $	\mathbf{Q}_\text{ss}\mathbf{o}\propto \mathbf{Q}_\text{ss} \mathbf{B}[ \boldsymbol{\delta}\mathbf{p}- \mathbf{p}_\text{ss}(\mathbf{p}_\text{ss}^\text{T}\mathbf{B}\boldsymbol{\delta}\mathbf{p})/ (\mathbf{p}_\text{ss}^\text{T}\mathbf{B}\mathbf{p}_\text{ss})] =\mathbf{Q}_\text{ss} \mathbf{B}[ \mathbf{p}- \mathbf{p}_\text{ss}(\mathbf{p}_\text{ss}^\text{T}\mathbf{B}\mathbf{p})/ (\mathbf{p}_\text{ss}^\text{T}\mathbf{B}\mathbf{p}_\text{ss})]$. Since $\mathbf{Q}_\text{ss}\mathbf{1}=\mathbf{0}$, this condition can also be expressed as
	\begin{equation}\label{eq:o}
		\mathbf{o}
		=\alpha \mathbf{B}\left( \mathbf{p}- \mathbf{p}_\text{ss}\frac{\mathbf{p}_\text{ss}^\text{T}\mathbf{B}\mathbf{p}}{ \mathbf{p}_\text{ss}^\text{T}\mathbf{B}\mathbf{p}_\text{ss}}\right)+\beta \mathbf{1},
	\end{equation}
	where $\alpha\neq0$. Then, for a given reference $\mathbf{p}$, the 
	bound in Eq.~\eqref{eq:mOUR_OUR} achieves its minimum for such observables, so that it is a valid bound for the uncertainty of any observables, while that minimum coincides with the bound in the corresponding OUR.
	\\
	
	\subsection{Towards tighter matrix bounds} \label{SM:mOUR_other}
	
	We now formally derive a fundamental bound from which the bounds in Sec.~\ref{SM:OUR_other} are shown to follow. Namely, [cf.~Eq.~(2) in the main text]
	\begin{equation}\label{eq:vOUR2}
		\lim_{t\,\rightarrow \infty} \,\frac{1}{t}\,\Delta^2 \overline{O}(t)=\mathbf{o}^\text{T}\mathbf{C}\mathbf{o}\geq\mathbf{o}^\text{T}\mathbf{Q}_\text{ss}^\text{T}  \left(\mathbf{B}-\boldsymbol{\delta}\mathbf{B}\right)^{-1}\! \mathbf{Q}_\text{ss}\mathbf{o},
	\end{equation}
	which is equivalent to the following matrix bound [cf.~Eq.~(3) in the main text],
	\begin{equation}\label{eq:mOUR2}
		\mathbf{C}\geq\mathbf{Q}_\text{ss}^\text{T}  \left(\mathbf{B}-\boldsymbol{\delta}\mathbf{B}\right)^{-1}\! \mathbf{Q}_\text{ss}.
	\end{equation}
	The bound in Eq.~\eqref{eq:mOUR2} is tighter than the mOUR as $\boldsymbol{\delta}\mathbf{B}$ defined in Eq.~\eqref{eq:deltaB} is positive semi-definite.	However, we only formally use the inverse $\left(\mathbf{B}-\boldsymbol{\delta}\mathbf{B}\right)^{-1}$ of $\mathbf{B}-\boldsymbol{\delta}\mathbf{B}$, which in contrast to $\mathbf{B}$  is not diagonal, cf.~Eqs.~\eqref{eq:B} and~\eqref{eq:deltaB}. In Sec.~\ref{SM:sat2}, we also show that the bound in Eq.~\eqref{eq:mOUR2} would saturate for any  cycle as well as equilibrium dynamics without asymptotic currents and thus finding the inverse of $\mathbf{B}-\boldsymbol{\delta}\mathbf{B}$ would be equivalent to calculating the variance rate for any system observable in those cases.\\
	
	\emph{Derivation of Eq.~\eqref{eq:vOUR2}}. From the derivation of Eq.~\eqref{eq:OUR2} in Sec.~\ref{SM:OUR_other}, we have [cf.~Eq.~\eqref{eq:OUR2_delta}]
	\begin{equation} \label{eq:OUR2_delta_inv}
		\mathbf{o}^\text{T}\mathbf{C}\mathbf{o}\geq \frac{(\boldsymbol{\delta}\mathbf{p}^\text{T}\mathbf{o})^2}{(\boldsymbol{\delta}\mathbf{p}-\delta\mathbf{p}_\text{ss})^\text{T}(\mathbf{B}-\boldsymbol{\delta}\mathbf{B})(\boldsymbol{\delta}\mathbf{p}-\delta \mathbf{p}_\text{ss} )},
	\end{equation} 
	where  $\boldsymbol{\delta}\mathbf{p}^\text{T}\mathbf{1}=0$. Then, introducing
	$\boldsymbol{\delta}\tilde{\mathbf{p}} \equiv \boldsymbol{\delta}\mathbf{p} -\delta\mathbf{p}_\text{ss}$,
	we have [cf.~Eq.~\eqref{eq:OUR_delta_inv_tilde}]
	\begin{equation}\label{eq:OUR2_delta_inv_tilde}
		\mathbf{o}^\text{T}\mathbf{C}\mathbf{o}\geq\frac{(\boldsymbol{\delta}{\tilde{\mathbf{p}}}^\text{T}\mathbf{Q}_\text{ss}\mathbf{o})^2}{\boldsymbol{\delta}{\tilde{\mathbf{p}}}^\text{T}(\mathbf{B}-\boldsymbol{\delta}\mathbf{B})\boldsymbol{\delta}\tilde{\mathbf{p}} }.
	\end{equation}
	Maximising the right-hand side with respect to $\boldsymbol{\delta}\tilde{\mathbf{p}}$, we directly obtain the bound in Eq.~\eqref{eq:vOUR2} for [cf.~Eq.~\eqref{eq:deltaptilde}] 
	\begin{equation}\label{eq:deltaptilde2}
		\boldsymbol{\delta}\tilde{\mathbf{p}}\propto(\mathbf{B}-\boldsymbol{\delta}\mathbf{B})^{-1} \mathbf{Q}_\text{ss} \mathbf{o},
	\end{equation}
	which can be seen by considering Cauchy-Schwarz inequality for the weighted scalar product $\langle \mathbf{u},\mathbf{v}\rangle\equiv\mathbf{u}^\text{T} (\mathbf{B}-\boldsymbol{\delta}\mathbf{B}) \mathbf{v}$, which is well defined since $\mathbf{B}-\mathbf{\delta B}>0$ [this follows from the strict positivity of the rate function $I( {o} )>0$ for $ {o} \neq \langle O\rangle_\text{ss}$, as $\mathbf{B}-\mathbf{\delta B}$ defines its upper bound in the leading order, cf.~Eqs.~\eqref{eq:l1OUR} and~\eqref{eq:OUR2_delta}]. The condition in Eq.~\eqref{eq:deltaptilde2} corresponds to  [cf.~Eq.~\eqref{eq:deltap}] 
	\begin{equation}\label{eq:deltap2}
		\boldsymbol{\delta}\mathbf{p}=\mathbf{Q}_\text{ss}^\text{T}\boldsymbol{\delta}\tilde{\mathbf{p}}\propto(\mathbf{B}-\boldsymbol{\delta}\mathbf{B})^{-1} \mathbf{Q}_\text{ss} \mathbf{o} - \left[\mathbf{1}^\text{T}(\mathbf{B}-\boldsymbol{\delta}\mathbf{B})^{-1}\mathbf{Q}_\text{ss} \mathbf{o}\right]\mathbf{p}_\text{ss}
	\end{equation}
	and $\delta=\mathbf{1}^\text{T}(\mathbf{B}-\boldsymbol{\delta}\mathbf{B})^{-1}\mathbf{Q}_\text{ss} \mathbf{o}$.
	\\
	
	\emph{Connections between Eqs.~\eqref{eq:OUR2} and~\eqref{eq:vOUR2}}.
	From the derivation above it follows that Eq.~\eqref{eq:vOUR2} can be formally obtained from Eq.~\eqref{eq:OUR2}, by considering a reference distribution $\mathbf{p}=\mathbf{p}_\text{ss}+\boldsymbol{\delta}\mathbf{p}$, where $\boldsymbol{\delta}\mathbf{p}$ is given by Eq.~\eqref{eq:deltap2}, so that [cf.~Eq.~\eqref{eq:p}]
	\begin{equation}\label{eq:p2}
		\mathbf{p}=\mathbf{p}_\text{ss}+\epsilon\left\{(\mathbf{B}-\boldsymbol{\delta}\mathbf{B})^{-1} \mathbf{Q}_\text{ss} \mathbf{o} - \left[\mathbf{1}^\text{T}(\mathbf{B}-\boldsymbol{\delta}\mathbf{B})^{-1}\mathbf{Q}_\text{ss} \mathbf{o}\right]\mathbf{p}_\text{ss} \right\}.
	\end{equation}
	This corresponds to a reference $	\mathbf{p}$ leading to the highest quality of the bound in Eq.~\eqref{eq:OUR2}  for a given observable $\mathbf{o}$.

	On the other hand, Eq.~\eqref{eq:OUR2}  is obtained by considering the  bound  on the uncertainty with respect to the reference $\mathbf{p}$ that follows from Eq.~\eqref{eq:vOUR2} [cf.~Eq.~\eqref{eq:mOUR_OUR}], 
	\begin{equation} \label{eq:mOUR2_OUR2}
		\frac{\mathbf{o}^\text{T}\mathbf{C}\mathbf{o}  } {[(\mathbf{p}-\mathbf{p}_\text{ss})^\text{T}\mathbf{o}]^2} \geq \frac{\mathbf{o}^\text{T}\mathbf{Q}_\text{ss}^\text{T}(\mathbf{B}-\boldsymbol{\delta}\mathbf{B})^{-1} \mathbf{Q}_\text{ss}\mathbf{o}} { [(\mathbf{p}-\mathbf{p}_\text{ss})^\text{T}\mathbf{o})]^2}.
	\end{equation}
	Indeed, an observable [cf.~Eq.~\eqref{eq:o}] 
	\begin{equation}\label{eq:o2}
		\mathbf{o}
		=\alpha (\mathbf{B}-\boldsymbol{\delta}\mathbf{B})\left[ \mathbf{p}- \mathbf{p}_\text{ss}\frac{\mathbf{p}_\text{ss}^\text{T}(\mathbf{B}-\boldsymbol{\delta}\mathbf{B})\mathbf{p}}{ \mathbf{p}_\text{ss}^\text{T}(\mathbf{B}-\boldsymbol{\delta}\mathbf{B})\mathbf{p}_\text{ss}}\right]+\beta \mathbf{1},
	\end{equation}
	where $\alpha\neq0$,
	leads to the lowest value of that bound, which is thus valid for all observables, whilst coinciding with that in Eq.~\eqref{eq:OUR2}.
	\\
	
	\section{Saturation of Occupation Uncertainty Relations and Matrix Occupation Uncertainty Relation}\label{SM:sat}
	
	We now present and derive conditions for the OUR for a given reference distribution, that is,  Eq.~(1) in the main text, to saturate for some system observable. The derivation also provides an alternative proof of this relation to that presented in Sec.~\ref{SM:OUR}. We further use those conditions to derive the condition for the mOUR for a given observable, i.e., Eq.~(2) in the main text, to saturate.  In turn, we arrive at the conditions for the mOUR,  Eq.~(3) in the main text, to saturate for all observables. Finally, we prove that those conditions are satisfied for the dynamics of unidirectional cycles and also discuss how it is not generally the case for equilibrium dynamics without asymptotic currents. Conditions for the saturation of the tighter bounds  in Eqs.~\eqref{eq:OUR2},~\eqref{eq:vOUR2}, and~\eqref{eq:mOUR2} derived in this Supplemental Material are uncovered as well.\\

	\emph{Relation to earlier work}.  We  now summarise the approach introduced in Ref.~\cite{Shiraishi2021} to investigate the asymptotic saturation of Thermodynamic Uncertainty Relation (TUR)~\cite{Barato2015,Gingrich2016} and Kinetic Uncertainty Relation (KUR)~\cite{Garrahan2017}. Below, we apply this approach to OURs.

	First, perturbing the dynamics generator from $\mathbf{W}$ for times $t<0$ to $\mathbf{W}+\epsilon\boldsymbol{\delta}\mathbf{W}$ for $t\geq 0$, the value of $\epsilon$  can be optimally estimated with respect to the signal to noise ratio by measuring a time-integrated dynamical observable  $\overline{O}_\text{opt}(t)$, where
	\begin{equation}\label{eq:O_opt_t}
		O_\text{opt}(t)=-\sum_{j}\delta_{j}(t)\sum_{k\neq j} (\boldsymbol{\delta}\mathbf{W})_{jk} +\sum_{j}\sum_{k\neq j}N_{jk}(t)\,\frac{(\boldsymbol{\delta}\mathbf{W})_{jk}}{(\mathbf{W})_{jk}}.
	\end{equation}
	Here, $\delta_j(t)=1$ if a configuration $j$ is occupied at time $t$ and $0$ otherwise (as in the main text), while
	$N_{jk}(t)=1$  if a transition from $j$ to $k$ occurs in the interval $[t,t+dt)$ and $0$ otherwise. When $t\rightarrow\infty$, $\overline{\delta}_{j}(t)/t\rightarrow \mu_{j}$ and $\overline{N}_{jk}(t)/t\rightarrow n_{jk}$, so that 
	\begin{equation}\label{eq:O_opt}
		\lim_{t\rightarrow\infty} \frac{1}{t}\overline{O}_\text{opt}(t)=-\sum_{j}\mu_{j}\sum_{k\neq j} (\boldsymbol{\delta}\mathbf{W})_{jk} +\sum_{j}\sum_{k\neq j}n_{jk}\,\frac{(\boldsymbol{\delta}\mathbf{W})_{jk}}{(\mathbf{W})_{jk}},
	\end{equation}
	while the large deviations of $\{\mu_j\}_j$ and $\{n_{jk}\}_{jk:\,j\neq k}$ are determined by the rate function in Eq.~\eqref{eq:l2.5}.
	
	Second, the maximal signal to ratio further equals the variance of the time-integrated dynamical observable, so that in the long time limit its rate
	\begin{equation}\label{eq:O_opt_var}
		\lim_{t\rightarrow\infty}\frac{1}{t}\Delta^2 \overline{O}_\text{opt}(t)=	\sum_{j}\sum_{k\neq j}(\mathbf{p}_\text{ss})_j\frac{[(\boldsymbol{\delta}\mathbf{W})_{jk}]^2}{(\mathbf{W})_{jk}}.
	\end{equation} 
	
	We note that the signal to noise ratio is unchanged when observables are rescaled by a (multiplicative) constant $\alpha$ or shifted by an (additive) constant $\beta$, so that the maximal signal to noise ratio will be also achieved for any $\overline{O}(t)=\alpha\overline{O}_\text{opt}(t) +\beta t$, where $\alpha\neq 0$.  Those will be the only time-integrated dynamical observables for which this occurs, as $\overline{O}_\text{opt}(t)$ corresponds to the logarithmic derivative for probability distribution for occupation times in individual configurations and total numbers of all possible transitions.  Similarly, in the long-time limit,  the rate in  Eq.~\eqref{eq:O_opt_var} is achieved if and only if are given by Eq.~\eqref{eq:O_opt} up to rescaling and shifting (and up to dependence of $\{\mu_j\}_j$ $\{n_{jk}\}_{jk: j\neq k}$, see the discussion below).

	Finally, it is known that perturbations of the generator can be chosen so that the inverse of the rate of the signal to noise ratio in the long-time limit becomes the rate of the uncertainty of currents or fluxes (cf.~Ref.~\cite{Terlizzi2019}). Thus, the uncertainty rate is bounded from above in terms of the corresponding optimal observable variance rate, cf.~Eq.~\eqref{eq:O_opt_var}. The latter rate  actually coincides with the so called minimal entropy production rate (which is bounded from above by the entropy production rate~\cite{Liu2020,Koyuk2020,Shiraishi2021}) or the activity. Thus, the asymptotic TUR or KUR is recovered.  Importantly,  those relations can then be saturated when the optimal observable in Eq.~\eqref{eq:O_opt} itself is an observable solely on currents or fluxes for which the TUR or KUR are valid (and, for TUR, the minimal entropy production rate coincides with the entropy production rate).

	\subsection{Conditions for saturating OURs}
	
	We show below that for a given reference distribution $\mathbf{p}$ the corresponding OUR , i.e., Eq.~(1) of the main text, can be saturated if and only if
	\begin{equation}\label{eq:OUR_cond}
		\sum_{j}\left(\sum_{k\neq j}	\,n_{jk}\right)\left[\frac{(\mathbf{p})_j}{(\mathbf{p}_\text{ss})_j}	-\frac{\mathbf{p}_\text{ss}^\text{T}\mathbf{B}\mathbf{p}}{ \mathbf{p}_\text{ss}^\text{T}\mathbf{B}\mathbf{p}_\text{ss}}\right] =0.
	\end{equation}
	When this condition is satisfied, the OUR saturates for observables
	\begin{equation}\label{eq:OUR_o_opt}
		\mathbf{o}=\alpha \mathbf{B}\left( \mathbf{p}- \mathbf{p}_\text{ss}\frac{\mathbf{p}_\text{ss}^\text{T}\mathbf{B}\mathbf{p}}{ \mathbf{p}_\text{ss}^\text{T}\mathbf{B}\mathbf{p}_\text{ss}}\right)+\beta \mathbf{1},
	\end{equation} 
	where $\alpha\neq0$. We note Eq.~\eqref{eq:OUR_o_opt} coincides with Eq.~\eqref{eq:o}. 
	\\
	
	\emph{Conditions for general dynamics}. Asymptotically, there are as many non-zero transition number rates as the non-zero transition rates, that, is $\mathbf{W}_{jk}=0$ implies  $n_{jk}= 0$. As the ergodicity is assumed, this number, which we denote $d'$, needs to obey
	\begin{equation}\label{eq:d'}
		d'\geq d.
	\end{equation}
	Indeed, all $d$ configuration need to be connected, which requires $d-1$ transition rates being non-zero, but those are unidirectional, so the stationary distribution would be supported on a single configuration only, and at least one more transition needs to be added for the stationary state to be supported everywhere.
	
	We also have so called continuity equations which asymptotically yield
	\begin{equation}\label{eq:n_jk }
		\sum_{k\neq j}	\,n_{jk}=\sum_{k\neq j}	\,n_{kj}
	\end{equation}
	for all $j$. These constitute $(d-1)$ conditions on the transition number rates $\{n_{jk}\}_{jk:\,j\neq k}$ [note that $\sum_j\sum_{k\neq j}	\,n_{jk}=\sum_k\sum_{j\neq k}n_{jk}$ which equals $\sum_j\sum_{k\neq j}	\,n_{kj}$ by swapping labels $j$ and $k$]. Thus, the transition number rates effectively form as a set of $d''$ random variables determining the level 2.5 large deviations~\cite{Maes2008,Bertini2015} [cf.~Eq.~\eqref{eq:l2.5}], where
	\begin{equation}\label{eq:d''}
		d''=d'-d+1. 
	\end{equation}
	Together with Eq.~\eqref{eq:d'}, we thus arrive at
	\begin{equation}\label{eq:d''2}
		d''\geq 1,
	\end{equation}
	with the equality for the unidirectional cycles. Generally
	$\mathbf{p}$ must be such that the condition in Eq.~\eqref{eq:OUR_cond} implies no further dependence between the remaining $d''$ random variables, i.e., should be fulfilled for any values attained by them. Below we focus on the cases of unidirectional cycles and equilibrium dynamics without asymptotic currents.\\
	
	\emph{Conditions for unidirectional cycles}. For unidirectional cycles, we have that [cf.~Eq.~\eqref{eq:d'}]
	\begin{equation}\label{eq:d'_uni}
		d'=d
	\end{equation}
	and thus [cf.~Eqs.~\eqref{eq:d''} and~\eqref{eq:d''2}]
	\begin{equation}\label{eq:d''_uni}
		d''=1.
	\end{equation}
	Eq.~\eqref{eq:d'_uni} and thus Eq.~\eqref{eq:d''_uni} are valid only for such dynamics. The single remaining variable due to Eq.~\eqref{eq:n_jk } is $n=n_{j (j+ 1)}$ when $(\mathbf{W})_{jk}=\lambda_j[-\delta_{jk}+\delta_{(j+1)k}]$ or  $n=n_{j (j-1)}$ when $(\mathbf{W})_{jk}=\lambda_j[-\delta_{jk}+\delta_{(j-1)k}]$. Therefore, Eq.~\eqref{eq:OUR_cond} becomes 
	\begin{equation}\label{eq:OUR_uni0}
		n	\left[\sum_{j}\frac{(\mathbf{p})_j}{(\mathbf{p}_\text{ss})_j}	-d\,\frac{\mathbf{p}_\text{ss}^\text{T}\mathbf{B}\mathbf{p}}{ \mathbf{p}_\text{ss}^\text{T}\mathbf{B}\mathbf{p}_\text{ss}}\right] =0,
	\end{equation}
	As $\mathbf{p}_\text{ss}=\mathbf{\Lambda}^{-1}\mathbf{1}/(\mathbf{1}^\text{T}\mathbf{\Lambda}^{-1}\mathbf{1})$, we then can express $\sum_{j}(\mathbf{p})_j/(\mathbf{p}_\text{ss})_j=\mathbf{1}^\text{T}\mathbf{\Lambda}\mathbf{p} (\mathbf{1}^\text{T}\mathbf{\Lambda}^{-1}\mathbf{1})$ while
	$\mathbf{p}_\text{ss}^\text{T}\mathbf{B}\mathbf{p}=\mathbf{1}^\text{T}\mathbf{\Lambda}\mathbf{p}$ and
	$\mathbf{p}_\text{ss}^\text{T}\mathbf{B}\mathbf{p}_\text{ss}=d/(\mathbf{1}^\text{T}\mathbf{\Lambda}^{-1}\mathbf{1})$. Therefore,
	\begin{equation}\label{eq:OUR_uni}
		\sum_{j}\frac{(\mathbf{p})_j}{(\mathbf{p}_\text{ss})_j}	-d\,\frac{\mathbf{p}_\text{ss}^\text{T}\mathbf{B}\mathbf{p}}{ \mathbf{p}_\text{ss}^\text{T}\mathbf{B}\mathbf{p}_\text{ss}}=0
	\end{equation}
	and the condition in Eq.~\eqref{eq:OUR_uni0} is satisfied. We note that this holds for any reference distribution $\mathbf{p}$. This further  leads to the saturation of the mOUR for unidirectional cycles, as we explain in the next section.\\

	\emph{Conditions for equilibrium dynamics}.  We focus on the case of equilibrium dynamics where the structure of transitions with non-zero rates ensures that no asymptotic currents can be present. In this case,
	\begin{equation}\label{eq:n_jk _eq}
		n_{jk}=n_{kj}
	\end{equation}
	for any $j$ and $k\neq j$, so that the continuity equations in Eq.~\eqref{eq:n_jk } are satisfied. Furthermore, 
	\begin{equation}\label{eq:d''_eq0}
		d''=\frac{d'}{2}.
	\end{equation}
	Indeed, for given $j$ and $k\neq j$, the transition rates are simultaneously $0$ or different from $0$ as on the average Eq.~\eqref{eq:n_jk _eq} yields $(\mathbf{p}_\text{ss})_j(\mathbf{W})_{jk}=(\mathbf{p}_\text{ss})_k(\mathbf{W})_{kj}$. Together with Eq.~\eqref{eq:d''}, this yields
	\begin{equation}\label{eq:d''_eq}
		d''=d-1.
	\end{equation}
	Furthermore, using Eq.~\eqref{eq:n_jk _eq} the condition in Eq.~\eqref{eq:OUR_cond} can be expressed as:
	\begin{equation}\label{eq:OUR_cond_eq0}
		\sum_{j}\sum_{k>j}	\,n_{jk}\left[\frac{(\mathbf{p})_j}{(\mathbf{p}_\text{ss})_j}+\frac{(\mathbf{p})_k}{(\mathbf{p}_\text{ss})_k}	-2\frac{\mathbf{p}_\text{ss}^\text{T}\mathbf{B}\mathbf{p}}{ \mathbf{p}_\text{ss}^\text{T}\mathbf{B}\mathbf{p}_\text{ss}}\right] =0.
	\end{equation}
	This condition states dependence between the remaining $(d-1)$ random variables and thus can  be satisfied if and only if
	\begin{equation}\label{eq:OUR_cond_eq}
		\frac{(\mathbf{p})_j}{(\mathbf{p}_\text{ss})_j}+\frac{(\mathbf{p})_k}{(\mathbf{p}_\text{ss})_k}	-2\frac{\mathbf{p}_\text{ss}^\text{T}\mathbf{B}\mathbf{p}}{ \mathbf{p}_\text{ss}^\text{T}\mathbf{B}\mathbf{p}_\text{ss}}=0
	\end{equation}
	[for all $j$ and $k\neq j$ such that $(\mathbf{W})_{jk}\neq 0$; there are $(d-1)$ such pairs from Eq.~\eqref{eq:d''_eq}]. 
	\\
	
	\emph{Derivation of Eqs.~\eqref{eq:OUR_cond} and~\eqref{eq:OUR_o_opt}}.
	For the expansions in Eq.~\eqref{eq:mu}, we can define the dynamics for which those rates become the average rates as
	\begin{equation}
		(\mathbf{W})_{jk}+\epsilon (\boldsymbol{\delta}	\mathbf{W})_{jk}+\mathcal{O}(\epsilon^2)=\frac{n_{jk}}{\mu_j}=\frac{	(1+\epsilon\delta)(\mathbf{W})_{jk}}{(\mathbf{p}_\text{ss})_j+\epsilon(\boldsymbol{\delta}\mathbf{p})_j}=(\mathbf{W})_{jk}-\epsilon (\mathbf{W})_{jk}\left[ \frac{(\boldsymbol{\delta}\mathbf{p})_j	}{(\mathbf{p}_\text{ss})_j}-\delta\right]+\mathcal{O}(\epsilon^2)
	\end{equation}
	for $j\neq k$,
	so that 
	\begin{equation}\label{eq:deltaW}
		(\boldsymbol{\delta}\mathbf{W})_{jk}=-(\mathbf{W})_{jk}\left[ \frac{(\boldsymbol{\delta}\mathbf{p})_j	}{(\mathbf{p}_\text{ss})_j}-\delta\right],
	\end{equation}
	including $j=k$ (so that $\boldsymbol{\delta}\mathbf{W} \mathbf{1}=\mathbf{0}$).
	Here, we have neglected the second and higher orders in $\epsilon$ as those do not contribute to the signal to noise ratio with respect to $\epsilon$ (cf.~Ref.~\cite{Liu2020}). 
	
	For the perturbation of the generator in Eq.~\eqref{eq:deltaW},  we have
	\begin{equation}
		\left(	\mathbf{p}_\text{ss}+ \epsilon	\boldsymbol{\delta}\mathbf{p}\right)^\text{T} \left (	\mathbf{W}+\epsilon \boldsymbol{\delta}	\mathbf{W}\right)=\mathbf{0}^\text{T}+\mathcal{O}(\epsilon^2),
	\end{equation}
	so that $\mathbf{p}_\text{ss}+ \epsilon	\boldsymbol{\delta}\mathbf{p}$ approximates  the stationary distribution of $	\mathbf{W}+\epsilon \boldsymbol{\delta}	\mathbf{W}$ up to the second order. It then follows that due to perturbing the generator, the rate for the average of the time-integrated observable, $\langle\overline{O}(t)\rangle/t$, asymptotically changes from  $\langle O\rangle_\text{ss}=\mathbf{p}_\text{ss}^\text{T}\mathbf{o}$ to $(\mathbf{p}_\text{ss}+\epsilon \boldsymbol{\delta}\mathbf{p})^\text{T}\mathbf{o}+\mathcal{O}(\epsilon^2)$. Thus, the rate of the signal from measuring that observable is $\boldsymbol{\delta}\mathbf{p}^\text{T}\mathbf{o}$, while the rate of its variance remains $\mathbf{o}^\text{T}\mathbf{C}\mathbf{o}$. 

	As the rate of the signal to noise ratio is bounded from above by Eq.~\eqref{eq:O_opt_var}, we then obtain
	\begin{equation} \label{eq:OUR_delta_O_opt}
		\frac{(\boldsymbol{\delta}\mathbf{p}^\text{T}\mathbf{o})^2}{\mathbf{o}^\text{T}\mathbf{C}\mathbf{o}}\leq \sum_{j}\sum_{k\neq j}(\mathbf{p}_\text{ss})_j\frac{[(\boldsymbol{\delta}\mathbf{W})_{jk}]^2}{(\mathbf{W})_{jk}}=\sum_{j}\sum_{k\neq j}(\mathbf{p}_\text{ss})_j(\mathbf{W})_{jk}\left[ \frac{(\boldsymbol{\delta}\mathbf{p})_j	}{(\mathbf{p}_\text{ss})_j}-\delta\right]^2=	(\boldsymbol{\delta}\mathbf{p}-\delta\mathbf{p}_\text{ss})^\text{T}\mathbf{B}( \boldsymbol{\delta}\mathbf{p}-\delta\mathbf{p}_\text{ss}).
	\end{equation}
	Here, we used the definition of $\mathbf{B}$ as in the main text [or, equivalently, in Eq.~\eqref{eq:B}]. 
	
	On the other hand, the optimal observable asymptotically corresponds to [cf.~Eq.~\eqref{eq:O_opt}]
	\begin{eqnarray}\label{eq:O_opt_OUR}
		\lim_{t\rightarrow\infty} \frac{1}{t}\overline{O}_\text{opt}(t)=-\sum_{j} \mu_{j}\left[ \frac{(\boldsymbol{\delta}\mathbf{p})_j	}{(\mathbf{p}_\text{ss})_j}-\delta\right]\sum_{k\neq j} (\mathbf{W})_{jk} \,+\sum_{j}\left(\sum_{k\neq j}	\,n_{jk}\right)\left[ \frac{(\boldsymbol{\delta}\mathbf{p})_j	}{(\mathbf{p}_\text{ss})_j}-\delta\right].
	\end{eqnarray}
	If
	\begin{equation}\label{eq:OUR_cond0}
		\sum_{j}\left(\sum_{k\neq j}	\,n_{jk}\right)\left[ \frac{(\boldsymbol{\delta}\mathbf{p})_j	}{(\mathbf{p}_\text{ss})_j}-\delta\right] =0,
	\end{equation}
	the optimal observable in Eq.~\eqref{eq:O_opt_OUR} is a system observable. Moreover,  the bound in Eq.~\eqref{eq:OUR_delta_O_opt} saturates for
	\begin{equation}\label{eq:OUR_o_opt0}
		(\mathbf{o})_j=\alpha\left[\frac{(\boldsymbol{\delta}\mathbf{p})_j	}{(\mathbf{p}_\text{ss})_j}-\delta\right]\sum_{k\neq j} (\mathbf{W})_{jk}+\beta.
	\end{equation}
	
	Finally, we note that the left-hand-side in Eq.~\eqref{eq:OUR_delta_O_opt}, is independent from $\delta$, so that only for $\delta$ leading to the minimum of the right-hand side, the bound may be saturated. This occurs for $\delta$ in Eq.~\eqref{eq:delta}. For $\mathbf{p}=\boldsymbol{\delta}\mathbf{p}+\mathbf{p}_\text{ss}$, Eqs.~\eqref{eq:OUR_cond0} and~\eqref{eq:OUR_o_opt0} then become Eqs.~\eqref{eq:OUR_cond} and~\eqref{eq:OUR_o_opt}. Crucially, in this case, the bound in Eq.~\eqref{eq:OUR_delta_O_opt} also coincides with the OUR  [cf.~Eqs.~\eqref{eq:OUR_delta} and~\eqref{eq:OUR_delta_opt}], so that the condition in Eq.~\eqref{eq:OUR_cond} becomes the condition for the saturation of OUR which occurs for observables as in Eq.~\eqref{eq:OUR_o_opt}. Therefore, this derivation of OUR can be seen as the alternative proof of Eq.~(1) in the main text to that in Sec.~\ref{SM:OUR_proof}.

	\subsection{Conditions for saturating mOUR}

	\emph{Conditions for saturation of mOUR for a given observable}.  We show below that for a  given observable $\mathbf{o}$, the corresponding mOUR, i.e., Eq.~(2) of the main text, saturates if and only if
	\begin{equation}\label{eq:vOUR_cond}
		\sum_{j}\left(\sum_{k\neq j}	\,n_{jk}\right) \frac{(\mathbf{o})_j- \mathbf{p}_\text{ss}^\text{T}\mathbf{o}}{\sum_{k\neq j} (\mathbf{W})_{jk}}  =0.
	\end{equation}\\

	\emph{Conditions for saturation of mOUR for all observables}.  Eq.~(3) of the main text saturates if Eq.~\eqref{eq:vOUR_cond} is fulfilled for all observables. This in general provides $d-1$ conditions, which should be fulfilled by $d''$ random variables that determine transition rates asymptotically [cf.~Eqs.~\eqref{eq:d'}-\eqref{eq:d''2}]. In particular, the conditions in Eq.~\eqref{eq:vOUR_cond} need to hold for observables corresponding to occupation times, which for configuration $j$ yields  
	\begin{equation}\label{eq:mOUR_cond}
		\left(\sum_{l\neq j}	\,n_{jl}\right) \frac{1}{\sum_{l\neq j} (\mathbf{W})_{jl}} - \sum_{k} 	\left(\sum_{l\neq k}	\,n_{kl}\right) \frac{ (\mathbf{p}_\text{ss})_j}{\sum_{l\neq k} (\mathbf{W})_{kl}} =0.
	\end{equation}
	We note that the condition  in Eq.~\eqref{eq:vOUR_cond} is recovered by multiplying  Eq.~\eqref{eq:mOUR_cond} by $(\mathbf{o})_j$ and summing over $j$ due to the linearity of the former in $\mathbf{o}$. Thus,
	Eq.~\eqref{eq:vOUR_cond} for all $j$ yields necessary and sufficient conditions for the saturation of the mOUR for all observables.\\
	
	\emph{Unidirectional cycles}. We now prove that the bound in Eq.~(3) of the main text saturates for unidirectional cycles. We present three approaches to achieve this. First, we show that Eq.~\eqref{eq:vOUR_cond} is indeed fulfilled for all observables. Second, we exploit the earlier shown fact that the OUR saturates for any reference distribution. Third, we revisit the derivation of Eqs.~(1) and~(3) presented in the Secs.~\ref{SM:OUR} and~\ref{SM:mOUR} of this Supplemental Material. \\ 
	
	\emph{First proof of saturation of mOUR for unidirectional cycles}. For  unidirectional cycles there is a single random variable 
	that determines transition rates asymptotically [cf.~Eq.~\eqref{eq:d''_uni}] 
	which is $n=n_{j (j+ 1)}$ when $(\mathbf{W})_{jk}=\lambda_j[-\delta_{jk}+\delta_{(j+1)k}]$ or  $n=n_{j (j-1)}$ when $(\mathbf{W})_{jk}=\lambda_j[-\delta_{jk}+\delta_{(j-1)k}]$ that coincides for all $j$. Then, in the former case, Eq.~\eqref{eq:vOUR_cond} reduces to 
	\begin{equation}\label{eq:vOUR_cond_uni}
		n	\sum_{j} \frac{(\mathbf{o})_j- \mathbf{p}_\text{ss}^\text{T}\mathbf{o}}{\sum_{k\neq j} (\mathbf{W})_{jk}}  =0.
	\end{equation}
	This is fulfilled as 
	\begin{equation}\label{eq:vOUR_cond_uni2}
		\sum_{j} \frac{(\mathbf{o})_j- \mathbf{p}_\text{ss}^\text{T}\mathbf{o}}{\sum_{k\neq j} (\mathbf{W})_{jk}}  =0,
	\end{equation}
	thanks to
	$(\mathbf{p}_\text{ss})_j=[1/\sum_{k\neq j} (\mathbf{W})_{jk}]/[\sum_{k} 1/\sum_{l\neq k} (\mathbf{W})_{kl}]$. Alternatively, Eq.~\eqref{eq:mOUR_cond} can be seen to be directly fulfilled due to this fact as it reduces to \begin{equation}\label{eq:mOUR_cond_uni}
		n\left[ \frac{1}{\sum_{l\neq j} (\mathbf{W})_{jl}} - \sum_{k}  \frac{ (\mathbf{p}_\text{ss})_j}{\sum_{l\neq k} (\mathbf{W})_{kl}} \right]=0.
	\end{equation}
	\\
	
	\emph{Second proof of saturation of mOUR for unidirectional cycles}. Since the condition for saturation of the OUR in  Eq.~\eqref{eq:OUR_cond} is fulfilled for any reference distribution [cf.~Eqs.~\eqref{eq:OUR_uni0} and~\eqref{eq:OUR_uni}], it is also valid satisfied for $\mathbf{p}$ in Eq.~\eqref{eq:p}, while Eq.~(2) of the main text is recovered and thus saturated [cf.~Eq.~\eqref{eq:mOUR_OUR}]. This occurs for any observable $\mathbf{o}$, and thus the mOUR saturates as well.\\
	
	\emph{Third proof of saturation of mOUR for unidirectional cycles}. Since for unidirectional  cycles only a single transition number rate $n$ appears in the level 2.5 rate function, this rate is then perturbatively optimised in the rate function bound used in derivation of Eq.~(1) of the main text. Then, the optimal choice of reference in Eq.~(2) of the main text can be seen as perturbatively optimising occupation times $\mu_j$ for a given value of the observable $ {o} $. Therefore, by the definition of the observable rate function $I( {o} )$ in Eq.~\eqref{eq:l1}, the perturbative bound on the observable rate function in Eq.~\eqref{eq:mOUR_full} then saturates in the leading order, and so does Eq.~(2) of the main text. This occurs for any observable $\mathbf{o}$, and thus the mOUR in Eq.~(3) of the main text saturates as well.\\

	\emph{Equilibrium dynamics}.  We now turn to equilibrium dynamics without asymptotic currents so that the number of fluxes is reduced by $2$ asymptotically [cf.~Eq.~\eqref{eq:d''_eq0}]. In that case, the condition for the saturation of Eq.~(2) of the main text given in  Eq.~\eqref{eq:vOUR_cond}  reduces to [cf.~Eq.~\eqref{eq:n_jk _eq}]
	\begin{equation}\label{eq:vOUR_cond_eq}
		\sum_{j}\sum_{k> j}	\,n_{jk}\left[\frac{(\mathbf{o})_j- \mathbf{p}_\text{ss}^\text{T}\mathbf{o}}{\sum_{l\neq j}  (\mathbf{W})_{jl}}  +\frac{(\mathbf{o})_k- \mathbf{p}_\text{ss}^\text{T}\mathbf{o}}{\sum_{l\neq k}  (\mathbf{W})_{kl}} \right]=0.
	\end{equation}
	This condition states dependence between remaining $d-1$ random variables  for fluxes [cf.~Eq.~\eqref{eq:d''_eq}], so it is fulfilled if and only if
	\begin{equation}\label{eq:vOUR_cond_eq2}
		\frac{(\mathbf{o})_j- \mathbf{p}_\text{ss}^\text{T}\mathbf{o}}{\sum_{l\neq j}  (\mathbf{W})_{jl}}  +\frac{(\mathbf{o})_k- \mathbf{p}_\text{ss}^\text{T}\mathbf{o}}{\sum_{l\neq k}  (\mathbf{W})_{kl}} =0
	\end{equation}
	for all $j$ and $k\neq j$ such that $(\mathbf{W})_{jk}\neq 0$; there are $(d-1)$ such pairs. 
	
	If the conditions in Eq.~\eqref{eq:vOUR_cond_eq2} are fulfilled for all observables, the mOUR in Eq.~(3)
	of the main text saturates.  Equivalently, the conditions need to hold for  observables corresponding to occupation times and thus  for each $j$ [cf.~Eq.~\eqref{eq:mOUR_cond}]
	\begin{equation}\label{eq:mOUR_cond_eq}
		\frac{1}{\sum_{l\neq j}  (\mathbf{W})_{jl}}  -\sum_k\frac{( \mathbf{p}_\text{ss})_j}{\sum_{l\neq k}  (\mathbf{W})_{kl}} =0.
	\end{equation}
	is required.
	This fixes the stationary distribution to be proportional to the inverse decay rates,  
	\begin{equation}\label{eq:mOUR_cond_eq2}
		(\mathbf{p}_\text{ss})_j\propto \frac{1}{\sum_{l\neq j}  (\mathbf{W})_{jl}} ,
	\end{equation}
	with the proportionality constant $\sum_k 1/\sum_{l\neq k}  (\mathbf{W})_{kl}$ from the normalisation $\mathbf{p}_\text{ss}^\text{T}\mathbf{1}=1$ (exactly as is the case for the cycles considered in Example I of the main text).
	
	Equation~\eqref{eq:mOUR_cond_eq2} holds for $d=2$, but it cannot be fulfilled when $d\geq 3$, as we now explain.  For a pair of configurations $j$ and $k$  such that $(\mathbf{W})_{jk}\neq 0$, we have $	(\mathbf{p}_\text{ss})_j/(\mathbf{p}_\text{ss})_k=(\mathbf{W})_{kj}/(\mathbf{W})_{jk}$. On the other hand, for ergodicity either $j$ or $k$ need to be connected to another configuration $l\neq j,k$. Since there are only $d-1$ connected pairs, there will be at least one configuration $j$ connected only to one another configuration $k$. Then, $j$ decay rate coincides with the transition rate to $k$, $\sum_{l\neq j}(\mathbf{W})_{jl}=(\mathbf{W})_{jk}$, but for $k$ those are different and $\sum_{l\neq k}(\mathbf{W})_{kl}>(\mathbf{W})_{kj}$. Thus, $(\mathbf{p}_\text{ss})_j/(\mathbf{p}_\text{ss})_k<[\sum_{l\neq k} (\mathbf{W})_{kl}]/[\sum_{l\neq j}(\mathbf{W})_{jl}]$, so the condition in Eq.~\eqref{eq:mOUR_cond_eq2} is not satisfied.\\

	\emph{Derivation of Eq.~\eqref{eq:vOUR_cond}}. When for a given observable $\mathbf{o}$, the corresponding mOUR, i.e., Eq.~(2) of the main text, saturates, so does the OUR for the best reference distribution chosen as in Eq.~\eqref{eq:p} (see Sec.~\ref{SM:mOUR_proof}). Thus, the condition in Eq.~\eqref{eq:OUR_cond} must be satisfied for $\mathbf{p}$ in Eq.~\eqref{eq:p}. In that case,
	\begin{equation}
		\frac{(\mathbf{p})_j}{(\mathbf{p}_\text{ss})_j}	-\frac{\mathbf{p}_\text{ss}^\text{T}\mathbf{B}\mathbf{p}}{ \mathbf{p}_\text{ss}^\text{T}\mathbf{B}\mathbf{p}_\text{ss}}=\epsilon \left[\frac{\left(\mathbf{B}^{-1} \mathbf{Q}_\text{ss} \mathbf{o} \right)_j}{(\mathbf{p}_\text{ss})_j}	-\frac{\mathbf{p}_\text{ss}^\text{T}\mathbf{B}\mathbf{B}^{-1} \mathbf{Q}_\text{ss} \mathbf{o} }{ \mathbf{p}_\text{ss}^\text{T}\mathbf{B}\mathbf{p}_\text{ss}}\right]=\epsilon \frac{\left(\mathbf{B}^{-1} \mathbf{Q}_\text{ss} \mathbf{o} \right)_j}{(\mathbf{p}_\text{ss})_j}=\epsilon\frac{(\mathbf{o})_j- \mathbf{p}_\text{ss}^\text{T}\mathbf{o}}{\sum_{k\neq j} (\mathbf{W})_{jk}},
	\end{equation}
	where we have used $\mathbf{p}_\text{ss}^\text{T}\mathbf{Q}_\text{ss}=\mathbf{0}$. This yields Eq.~\eqref{eq:vOUR_cond}.

	\subsection{Conditions for saturating tighter bounds}\label{SM:sat2}
	
	\emph{Conditions for saturation of tighter bounds}. Below we show that for a given reference distribution $\mathbf{p}$, the corresponding tighter bound in Eq.~\eqref{eq:OUR2} can be saturated if and only if
	\begin{equation}\label{eq:OUR2_cond}
		\sum_{j}\sum_{k\neq j}	\,n_{jk}\frac{(\mathbf{p})_j(\mathbf{W})_{jk}-(\mathbf{p})_k(\mathbf{W})_{kj}-\frac{\mathbf{p}_\text{ss}^\text{T}(\mathbf{B}-\boldsymbol{\delta}\mathbf{B})\mathbf{p}}{ \mathbf{p}_\text{ss}^\text{T}(\mathbf{B}-\boldsymbol{\delta}\mathbf{B})\mathbf{p}_\text{ss}} \left[(\mathbf{p}_\text{ss})_j(\mathbf{W})_{jk}-(\mathbf{p}_\text{ss})_k(\mathbf{W})_{kj}\right]}{ (\mathbf{p}_\text{ss})_j(\mathbf{W})_{jk}+(\mathbf{p}_\text{ss})_k(\mathbf{W})_{kj}} =0.
	\end{equation}
	When this condition is satisfied, the bound saturates for observables
	\begin{equation}\label{eq:OUR2_o_opt}
		\mathbf{o}=\alpha (\mathbf{B}-\boldsymbol{\delta}\mathbf{B})\left[ \mathbf{p}- \mathbf{p}_\text{ss}\frac{\mathbf{p}_\text{ss}^\text{T}(\mathbf{B}-\boldsymbol{\delta}\mathbf{B})\mathbf{p}}{ \mathbf{p}_\text{ss}^\text{T}(\mathbf{B}-\boldsymbol{\delta}\mathbf{B})\mathbf{p}_\text{ss}}\right]+\beta \mathbf{1},
	\end{equation} 
	where $\alpha\neq0$. We note Eq.~\eqref{eq:OUR2_o_opt} coincides with Eq.~\eqref{eq:o2}.
	
	It follows that that for a given observable $\mathbf{o}$, the corresponding tighter matrix bound in Eq.~\eqref{eq:vOUR2} saturates if and only if [cf.~Eq.~\eqref{eq:vOUR_cond}]
	\begin{equation}\label{eq:vOUR2_cond}
		\sum_{j}\left(\sum_{k\neq j}	\,n_{jk}\right) \frac{[(\mathbf{B}-\boldsymbol{\delta}\mathbf{B})^{-1}\mathbf{Q}_\text{ss}\mathbf{o}]_j(\mathbf{W})_{jk}-[(\mathbf{B}-\boldsymbol{\delta}\mathbf{B})^{-1}\mathbf{Q}_\text{ss}\mathbf{o}]_k(\mathbf{W})_{kj}}{ (\mathbf{p}_\text{ss})_j(\mathbf{W})_{jk}+(\mathbf{p}_\text{ss})_k(\mathbf{W})_{kj}}  =0.
	\end{equation}
	Finally, the tighter matrix bound  for all observables, that is, Eq.~\eqref{eq:mOUR2} saturates if and only if for each $l$ [cf.~Eq.~\eqref{eq:mOUR_cond}]
	\begin{eqnarray}\nonumber
		&&\sum_{j}\!\!\left(\sum_{k\neq j}	n_{jk}\!\right)\!\! \frac{[(\mathbf{B}\!-\!\boldsymbol{\delta}\mathbf{B})^{-1}]_{jl}(\mathbf{W})_{jk}-[(\mathbf{B}\!-\!\boldsymbol{\delta}\mathbf{B})^{-1}]_{kl}(\mathbf{W})_{kj} -\{[(\mathbf{B}\!-\!\boldsymbol{\delta}\mathbf{B})^{-1}\mathbf{1}]_{j} (\mathbf{W})_{jk}-[(\mathbf{B}\!-\!\boldsymbol{\delta}\mathbf{B})^{-1}\mathbf{1}]_{k} (\mathbf{W})_{kj}\}(\mathbf{p}_\text{ss})_{l}}{ (\mathbf{p}_\text{ss})_j(\mathbf{W})_{jk}+(\mathbf{p}_\text{ss})_k(\mathbf{W})_{kj}}  
		\\\label{eq:mOUR2_cond}&&=0.
	\end{eqnarray}
	\\
	
	\emph{General dynamics}. Since the bounds in Eqs.~\eqref{eq:OUR2},~\eqref{eq:vOUR2}, and~\eqref{eq:mOUR2} are tighter than the bounds in Eqs. (1), (2), and (3) of the main text, the conditions in Eqs.~\eqref{eq:OUR2_cond}, \eqref{eq:vOUR2_cond}, and~\eqref{eq:mOUR2_cond} must be less restrictive than those in  Eqs.~\eqref{eq:OUR_cond}, \eqref{eq:vOUR_cond}, and~\eqref{eq:mOUR_cond}, respectively. Indeed, below we argue that Eq.~\eqref{eq:OUR2_cond} is satisfied both  for any cycle as well as for equilibrium dynamics where no asymptotic currents can be present even if transitions rates are changed, so that the tighter bound in Eq.~\eqref{eq:OUR2} can saturate for any choice of reference distribution $\mathbf{p}$. It then follows that the tighter matrix bounds in Eqs.~\eqref{eq:vOUR2} and~\eqref{eq:mOUR2} both saturate. \\ 
	
	\emph{Cycles}.  For any unidirectional cycle [cf.~Eq.~\eqref{eq:deltaB}]
	\begin{equation}
		\delta \mathbf{B}=\mathbf{0}.
	\end{equation} 
	Thus, the condition in Eq.~\eqref{eq:OUR2_cond} reduces to that in Eq.~\eqref{eq:OUR_cond}, which has already been shown to hold for such dynamics. This is as expected, since unidirectional cycles feature only a single flux asymptotically, and thus derivations of Eq.~(1) in the main text and of Eq.~\eqref{eq:OUR2} are equivalent (see Secs.~\ref{SM:OUR_proof} and~\ref{SM:OUR_other}). For a general  cycle, however, this is no longer the case, as up to $d-1$ symmetric fluxes [$n_{j(j+1)}+n_{(j+1)j}$] are present asymptotically together with a single asymmetric flux [$n_{j(j+1)}-n_{(j+1)j}$ which is the same for all $j$ due to the stationarity conditions in Eq.~\eqref{eq:n_jk }]. Nevertheless, all those fluxes are optimised in the derivation of Eq.~\eqref{eq:vOUR2}, while the optimal reference choice in Eq.~\eqref{eq:p2} correspond to optimising occupation times or a given value of the observable, so that the contraction principle in Eq.~\eqref{eq:l1} is realised perturbatively. Therefore,  Eq.~\eqref{eq:vOUR2} saturates for all cycles. \\

	\emph{Equilibrium dynamics}. For any equilibrium dynamics, the detailed balance conditions holds with $(\mathbf{p}_\text{ss})_j(\mathbf{W})_{jk}=(\mathbf{p}_\text{ss})_k(\mathbf{W})_{kj}$ ensure, so that  Eq.~\eqref{eq:OUR2_cond}  reduces to 
	\begin{equation}\label{eq:OUR2_cond_eq}
		\sum_{j}\sum_{k> j}	\left(n_{jk}-n_{kj}\right)\frac{(\mathbf{p})_j(\mathbf{W})_{jk}-(\mathbf{p})_k(\mathbf{W})_{kj}}{ (\mathbf{p}_\text{ss})_j(\mathbf{W})_{jk}+(\mathbf{p}_\text{ss})_k(\mathbf{W})_{kj}} =0.
	\end{equation}
	If no currents are present asymptotically, Eq.~\eqref{eq:n_jk _eq} holds,  and hence the above condition is satisfied for any $\mathbf{p}$.  Again, as this leads to Eq.~\eqref{eq:vOUR2} also saturating, it can be also seen as a consequence of the rate function fully contracted in the derivation of Eqs.~\eqref{eq:vOUR2}. Indeed, as no currents are present asymptotically, all fluxes and occupation times are optimised.
	\\

	\emph{Derivation of Eqs.~\eqref{eq:OUR2_cond} and~\eqref{eq:OUR2_o_opt}}.
	For the expansions in Eq.~\eqref{eq:mu2},  the dynamics for which those rates become the average rates is
	\begin{equation}
		(\mathbf{W})_{jk}+\epsilon (\boldsymbol{\delta}	\mathbf{W})_{jk}+\mathcal{O}(\epsilon^2)=\frac{n_{jk}}{\mu_j}=\frac{	(1+\epsilon\delta)(\mathbf{W})_{jk}+\epsilon \delta_{jk}}{(\mathbf{p}_\text{ss})_j+\epsilon(\boldsymbol{\delta}\mathbf{p})_j}=(\mathbf{W})_{jk}-\epsilon (\mathbf{W})_{jk}\left[ \frac{(\boldsymbol{\delta}\mathbf{p})_j	}{(\mathbf{p}_\text{ss})_j}-\delta-\frac{\delta_{jk}}{(\mathbf{p}_\text{ss})_j(\mathbf{W})_{jk}} \right]+\mathcal{O}(\epsilon^2),
	\end{equation}
	where $j\neq k$ and $\delta_{jk}=\delta_{kj}$,
	so that in the linear order 
	\begin{equation}\label{eq:deltaW2}
		(\boldsymbol{\delta}\mathbf{W})_{jj}=-(\mathbf{W})_{jj}\left[ \frac{(\boldsymbol{\delta}\mathbf{p})_j	}{(\mathbf{p}_\text{ss})_j}-\delta\right]-\sum_{k\neq j}\frac{\delta_{jk}}{(\mathbf{p}_\text{ss})_j}, \quad 	(\boldsymbol{\delta}\mathbf{W})_{jk}=-(\mathbf{W})_{jk}\left[ \frac{(\boldsymbol{\delta}\mathbf{p})_j	}{(\mathbf{p}_\text{ss})_j}-\delta-\frac{\delta_{jk}}{(\mathbf{p}_\text{ss})_j(\mathbf{W})_{jk}}\right] \ \text{for}\ j\neq k.
	\end{equation}
	We then have  that $\mathbf{p}_\text{ss}+ \epsilon	\boldsymbol{\delta}\mathbf{p}$ approximates  the stationary distribution of $	\mathbf{W}+\epsilon \boldsymbol{\delta}	\mathbf{W}$ as
	\begin{equation}
		\left(	\mathbf{p}_\text{ss}+ \epsilon	\boldsymbol{\delta}\mathbf{p}\right)^\text{T} \left (	\mathbf{W}+\epsilon \boldsymbol{\delta}	\mathbf{W}\right)=\mathbf{0}^\text{T}+\mathcal{O}(\epsilon^2),
	\end{equation}
	so that the rate of the signal from measuring  observable $\overline{O}(t)$ is $\boldsymbol{\delta}\mathbf{p}^\text{T}\mathbf{o}$, while the rate of its variance remains $\mathbf{o}^\text{T}\mathbf{C}\mathbf{o}$. 
	From Eq.~\eqref{eq:O_opt_var}, 
	\begin{equation} \label{eq:OUR2_delta_O_opt}
		\frac{(\boldsymbol{\delta}\mathbf{p}^\text{T}\mathbf{o})^2}{\mathbf{o}^\text{T}\mathbf{C}\mathbf{o}}\leq \sum_{j}\sum_{k\neq j}(\mathbf{p}_\text{ss})_j\frac{[(\boldsymbol{\delta}\mathbf{W})_{jk}]^2}{(\mathbf{W})_{jk}}=\sum_{j}\sum_{k\neq j}(\mathbf{p}_\text{ss})_j(\mathbf{W})_{jk}\left[ \frac{(\boldsymbol{\delta}\mathbf{p})_j	}{(\mathbf{p}_\text{ss})_j}-\delta-\frac{\delta_{jk}}{(\mathbf{p}_\text{ss})_j(\mathbf{W})_{jk}}\right]^2,
	\end{equation}
	while the optimal observable asymptotically corresponds to [cf.~Eq.~\eqref{eq:O_opt}]
	\begin{eqnarray}\label{eq:O_opt_OUR2}
		\lim_{t\rightarrow\infty} \frac{1}{t}\overline{O}_\text{opt}(t)=-\sum_{j} \mu_{j}\left\{\left[ \frac{(\boldsymbol{\delta}\mathbf{p})_j	}{(\mathbf{p}_\text{ss})_j}-\delta\right]\sum_{k\neq j} (\mathbf{W})_{jk}-\sum_{k\neq j} \frac{\delta_{jk}}{(\mathbf{p}_\text{ss})_j}\right\} +\sum_{j}\sum_{k\neq j}	\,n_{jk}\left[ \frac{(\boldsymbol{\delta}\mathbf{p})_j	}{(\mathbf{p}_\text{ss})_j}-\delta-\frac{\delta_{jk}}{(\mathbf{p}_\text{ss})_j(\mathbf{W})_{jk}}\right].
	\end{eqnarray}
	If
	\begin{equation}\label{eq:OUR2_cond0}
		\sum_{j}\sum_{k\neq j}	\,n_{jk}\left[ \frac{(\boldsymbol{\delta}\mathbf{p})_j	}{(\mathbf{p}_\text{ss})_j}-\delta-\frac{\delta_{jk}}{(\mathbf{p}_\text{ss})_j(\mathbf{W})_{jk}}\right] =0,
	\end{equation}
	the optimal observable in Eq.~\eqref{eq:O_opt_OUR2} is a system observable. Moreover,  the bound in Eq.~\eqref{eq:OUR2_delta_O_opt} saturates for
	\begin{equation}\label{eq:OUR2_o_opt0}
		(\mathbf{o})_j=\alpha\left\{\left[\frac{(\boldsymbol{\delta}\mathbf{p})_j	}{(\mathbf{p}_\text{ss})_j}-\delta\right]\sum_{k\neq j} (\mathbf{W})_{jk}-\sum_{k\neq j} \frac{\delta_{jk}}{(\mathbf{p}_\text{ss})_j} \right\}+\beta.
	\end{equation}
	
	The left-hand-side in Eq.~\eqref{eq:OUR_delta_O_opt}, is independent from $\{\delta_{jk}\}_{jk:\,j>k}$ and $\delta$, so that only for the minimum of the right-hand side, the bound may be saturated. This occurs when Eqs.~\eqref{eq:delta_jk} and~\eqref{eq:delta2} hold. For $\mathbf{p}=\boldsymbol{\delta}\mathbf{p}+\mathbf{p}_\text{ss}$, Eqs.~\eqref{eq:OUR2_cond0} and~\eqref{eq:OUR2_o_opt0} then become Eqs.~\eqref{eq:OUR2_cond} and~\eqref{eq:OUR2_o_opt}. Crucially, in this case, the bound in Eq.~\eqref{eq:OUR2_delta_O_opt} also coincides with the tighter bound in Eq.~\eqref{eq:OUR2}  [cf.~Eqs.~\eqref{eq:OUR2_delta0},~\eqref{eq:OUR2_delta}, and~\eqref{eq:OUR2_delta_opt}], so that the condition in Eq.~\eqref{eq:OUR2_cond} becomes the condition for its saturation which occurs for observables as in Eq.~\eqref{eq:OUR2_o_opt}. Therefore, this derivation  can be seen as the alternative proof of Eq.~\eqref{eq:OUR2} to that in Sec.~\ref{SM:OUR_other}. \\
	
	\emph{Derivation of Eq.~\eqref{eq:vOUR2_cond}}.
	When for a given observable $\mathbf{o}$, the bound in Eq.~\eqref{eq:vOUR2} saturates, so does the bound in Eq.~\eqref{eq:OUR2} for the best reference distribution chosen as in Eq.~\eqref{eq:p2} (see Sec.~\ref{SM:mOUR_other}). Thus, the condition in Eq.~\eqref{eq:OUR2_cond} must be satisfied for $\mathbf{p}$ in Eq.~\eqref{eq:p2}. In that case,
	\begin{eqnarray}
		\frac{(\mathbf{p})_j(\mathbf{W})_{jk}-(\mathbf{p})_k(\mathbf{W})_{kj}-\frac{\mathbf{p}_\text{ss}^\text{T}(\mathbf{B}-\boldsymbol{\delta}\mathbf{B})\mathbf{p}}{ \mathbf{p}_\text{ss}^\text{T}(\mathbf{B}-\boldsymbol{\delta}\mathbf{B})\mathbf{p}_\text{ss}} \left[(\mathbf{p}_\text{ss})_j(\mathbf{W})_{jk}-(\mathbf{p}_\text{ss})_k(\mathbf{W})_{kj}\right]}{ (\mathbf{p}_\text{ss})_j(\mathbf{W})_{jk}+(\mathbf{p}_\text{ss})_k(\mathbf{W})_{kj}}
		\\\nonumber
		=\epsilon\frac{[(\mathbf{B}-\boldsymbol{\delta}\mathbf{B})^{-1}\mathbf{Q}_\text{ss}\mathbf{o}]_j(\mathbf{W})_{jk}-[(\mathbf{B}-\boldsymbol{\delta}\mathbf{B})^{-1}\mathbf{Q}_\text{ss}\mathbf{o}]_k(\mathbf{W})_{kj}}{ (\mathbf{p}_\text{ss})_j(\mathbf{W})_{jk}+(\mathbf{p}_\text{ss})_k(\mathbf{W})_{kj}},
	\end{eqnarray}
	where we have used $\mathbf{p}_\text{ss}^\text{T}\mathbf{Q}_\text{ss}=\mathbf{0}$. This yields Eq.~\eqref{eq:vOUR_cond}. 
	\\
	
	\emph{Derivation of Eq.~\eqref{eq:mOUR2_cond}}.  Eq.~\eqref{eq:mOUR2} saturates if Eq.~\eqref{eq:vOUR2} holds for all observables and thus Eq.~\eqref{eq:vOUR2_cond} is satisfied for all observables.  Eq.~\eqref{eq:mOUR2_cond} corresponds to Eq.~\eqref{eq:vOUR2_cond} for the observable corresponding to the occupation time in a configuration $l$. Since by multiplying  Eq.~\eqref{eq:mOUR2_cond} by $(\mathbf{o})_l$ and summing over $l$, Eq.~\eqref{eq:vOUR2_cond} is recovered, Eq.~\eqref{eq:mOUR2_cond} for all $l$ yields necessary and sufficient conditions for the saturation of Eq.~\eqref{eq:mOUR2} for all observables.

	\section{Bounds in terms of L2 norms} \label{SM:L2}

	Here, we show how variances of the time integrals can be bounded in terms of squared L2 norm for the observable shifted by its average. An analogous result is also given for the variance in the stationary distribution.\\

	The rates of correlations between occupation times can be bounded as
	\begin{equation} \label{eq:mOUR_Q}
		\mathbf{C}
		\,
		\geq \min_j \frac{(\mathbf{p}_\text{ss})_{j}}{(\mathbf{\Lambda})_{j j}}\, \mathbf{Q}_\text{ss} ^\text{T}\mathbf{Q}_\text{ss} 
		\geq \frac{p_{\min}}{\lambda_{\max}} \mathbf{Q}_\text{ss} ^\text{T}\mathbf{Q}_\text{ss}. 
	\end{equation}
	where $p_{\min}\equiv\min_{j}(\mathbf{p}_\text{ss})_{j}$ is the minimum probability in the stationary distribution (assumed $>0$ in this work). 
	The first inequality follows from Eq.~(3)  in the main text since $\mathbf{B}^{-1}\geq \min_j [(\mathbf{p}_\text{ss})_{j}/(\mathbf{\Lambda})_{j j}] \mathbf{I}$ and the second by observing $\min_j [(\mathbf{p}_\text{ss})_{j}/(\mathbf{\Lambda})_{j j}]\geq {p_{\min}}/{\lambda_{\max}}$. The second inequality can also be obtained directly from Eq.~(4) in the main text as $\mathbf{D}\geq p_{\min}\mathbf{I}$, which leads to
	\begin{equation} \label{eq:mOUR_D}
		\mathbf{\Delta}^2
		\,
		\geq p_{\min} \mathbf{Q}_\text{ss} ^\text{T}\mathbf{Q}_\text{ss}. 
	\end{equation}

	The performance of the first bound in Eq.~\eqref{eq:mOUR_Q} and coincides with that of Eq.~(3) if the ratio of decay rates and the stationary distributions is uniform. The second bound coincides with the bound in Eq.~(3) if both the decay rates and the stationary distributions are uniform. That is, $(\mathbf{\Lambda})_{jj}=\lambda_{\max}$ and  $(\mathbf{p}_\text{ss})_{j}=p_{\min}$ for all $j$. Finally, the bound in  Eq.~\eqref{eq:mOUR_D} saturates if the distribution is uniform, and  $p_{\min}=1/d$.

\end{document}